\documentclass[aps,showpacs,twocolumn,pra,superscriptaddress]{revtex4}

\usepackage[tbtags]{amsmath}
\usepackage{amssymb}
\usepackage{amsfonts}
\usepackage{graphicx}
\usepackage{bmpsize}
\usepackage{mathtools}
\usepackage[colorlinks=true,citecolor=blue,urlcolor=blue]{hyperref}
\usepackage{appendix}

\hypersetup{colorlinks,linkcolor=blue,filecolor=blue,urlcolor=blue,citecolor=blue,}
\usepackage{multirow}

\begin{document}
	
	\title{Quantum Stirling heat engine based on Two-qubit Quantum Rabi Model with Spin-Spin Coupling}
	
	\author{Luxin Xu}
	\affiliation{Key Laboratory of Low-Dimension Quantum Structures and Quantum Control of Ministry of Education, Synergetic Innovation Center for Quantum Effects and Applications, Xiangjiang-Laboratory, Institute of Interdisciplinary Studies and Department of Physics, Hunan Normal University, Changsha 410081, China}

    \author{Chunfeng Wu}\thanks{Corresponding author: chunfeng\_wu@sutd.edu.sg}
    \affiliation{Science, Mathematics and Technology, Singapore University of Technology and Design, 8 Somapah Road, Singapore 487372, Singapore}

	\author{Changliang Ren}\thanks{Corresponding author: renchangliang@hunnu.edu.cn}
	\affiliation{Key Laboratory of Low-Dimension Quantum Structures and Quantum Control of Ministry of Education, Synergetic Innovation Center for Quantum Effects and Applications, Xiangjiang-Laboratory, Institute of Interdisciplinary Studies and Department of Physics, Hunan Normal University, Changsha 410081, China}
	
	\begin{abstract}

 Enhancing the efficiency of quantum heat engines (QHEs) is crucial for advancing fundamental research and quantum technology. {\color{black}We here we explore a quantum Stirling cycle using a two-qubit quantum Rabi model with spin-spin coupling as a working medium. We propose parameter optimization strategies to maximize the efficiency of the heat engine, as there are multiple ways for the effective coupling constant to move toward its critical value. In the normal phase of the system, the efficiency can be improved by increasing the temperature ratio of hot-to-cold reservoirs and enhancing spin-mode coupling strength. However, increasing spin-spin coupling strength inhibits the improvement of the efficiency. As the system goes to its critical point, QHE efficiency under low-temperature conditions tends to the Carnot limit. In the superradiant phase, the efficiency approaches the Carnot limit more closely as the cold reservoir's temperature decreases given a constant temperature ratio. Conversely, when the cold reservoir's temperature rises, the efficiency increases due to a higher ratio of spin-mode coupling strength to mode frequency.} If the spin-spin coupling strength is constant, increasing the hot-to-cold reservoir temperature ratio requires a corresponding increase in spin-mode coupling strength to achieve the Carnot efficiency. Our work deepens the understanding of QHE performance under various conditions and {\color{black} provides operative methods for optimizing the efficiency of QHE.}
\end{abstract}
\maketitle

\section{Introduction}

	Quantum thermodynamics \cite{scovil1959three,kosloff1984quantum, Kosloff2013Quantum, goold2016d, Sai2016Quantum,deffner2019quantum, quan2007quantum}, which combines the principles of quantum mechanics and thermodynamics, plays a crucial role in understanding the microscopic mechanisms of energy conversion and has become a prominent research area. {\color{black} As an important part of this field, quantum heat engines (QHEs) enhance our understanding of quantum thermodynamics \cite{quan2007quantum,scully2010quantum, PhysRevE.86.051105} and have practical significance in experiments and applications.} {\color{black} It has been shown in the literature that QHEs can achieve efficient heat-to-work conversion when quantum effects are inherently exhibited in a working medium or reservoir
  \cite{scully2010quantum, Scully2011QuantumHE, PhysRevLett.122.110601, Kammerlander2015CoherenceAM,PhysRevLett.119.170602, Dorfman2018EfficiencyAM,Dillenschneider_2009, Abah_2014,Us_natcom, PhysRevX.7.031044,PhysRevLett.112.030602,PhysRevE.93.052120, PhysRevE.86.051105,CF16, Fogarty_2021, PhysRevE.96.022143}, leading to substantially boost performance.}  {\color{black} Moreover, recent research studies have demonstrated the impact of different system interactions on the efficiency of }{\color{black} QHEs}, such as collective cooperative effects \cite{Niedenzu_2018},super-radiance \cite{hardal2015superradiant}, many-body localization \cite{PhysRevB.99.024203}, and phase transitions \cite{PhysRevE.98.042112, CF16, Fogarty_2021, PhysRevE.98.052124}. Experimentally, QHEs have been realized in various {\color{black} physical systems}, including trapped ions \cite{PhysRevLett.109.203006, PhysRevLett.112.030602}, ultracold atoms \cite{PhysRevLett.108.085303,doi:10.1126/science.1242308}, optomechanical systems \cite{PhysRevLett.112.150602}, NMR techniques \cite{PhysRevLett.123.240601}, nitrogen vacancy centers in diamonds \cite{PhysRevLett.122.110601}, superconducting circuits\cite{PhysRevApplied.17.064022,Ronzani2018TunablePH,pekola2019thermodynamics}\textit{etc}.
	
	{\color{black} A series of protocols for implementing quantum-phase-transition-assisted(QPT-assisted) QHEs have been proposed \cite{CF16, Fogarty_2021, PhysRevE.96.022143, PhysRevE.98.052147, PhysRevResearch.2.043247, purkait2022performance,alcalde2019quantum, PhysRevA.109.022208,chen2019interaction, PhysRevE.94.052122,kloc2019collective, Piccitto_2022,revathy2024quantum,bs2022bath,jussiau2023many}, and specifically, the efficiency of different heat engine cycles close to the critical point has been explored in the references.} Compared to {\color{black} the Carnot engine }\cite{Dann_2020,bender2000quantum,denzler2021power,sutantyo2020three,ccakmak2020construction,dann2020quantum,qiu2020quantum,qstac635abib41} and {\color{black} the Otto engine }\cite{kloc2019collective,CF16,Fogarty_2021,PhysRevE.98.052147,PhysRevResearch.2.043247} cycles, {\color{black} the investigations on critical quantum Stirling heat engines(QSHEs)} \cite{purkait2022performance, PhysRevLett.120.100601, PhysRevLett.106.070401, PhysRevE.98.052124} are relatively less extensive. Ma \textit{et al.} explored the {\color{black} the Stirling cycle using the Lipkin-Meshkov-Glick (LMG) model as the working medium. They showed that the quantum phase transition (QPT) of the LMG model at low temperatures enhances the efficiency of the {\color{black} QHE}, approaching Carnot efficiency \cite{PhysRevE.96.022143}.
 } Further studies indicated that the Stirling cycle's efficiency can be significantly improved when the anisotropic LMG model or Dicke model is used as {\color{black}a} working medium, even achieving the Carnot limit during {\color{black} a QPT in thermodynamic equilibrium} \cite{alcalde2019quantum}. However, the LMG model and the Dicke model only exhibit QPT in the thermodynamic limit when the number of particles or atoms approaches infinity, $N \to \infty$ \cite{alcalde2019quantum, PhysRevE.96.022143}. In systems with a finite number of particles, it is difficult to accurately capture the essential behavior of the QPTs.
	
In 2024, Wang \textit{et al.} used the quantum Rabi model (QRM) as {\color{black} a} working medium of the Stirling cycle and studied the effect of criticality on its efficiency \cite{PhysRevA.109.022208}. The {\color{black} QRM} is a typical {\color{black} finite-body} system that can exhibit QPT \cite{PhysRevLett.115.180404, PhysRevLett.118.073001, ying2020quantum, PhysRevLett.119.220601,fallas2022understanding} and has received widespread attention in recent years \cite{PhysRevLett.119.220601,ying2022critical,chen2021experimental, Larson_2017,PhysRevA.92.053823, Lu_2024}. The QPT of the QRM exists when the thermodynamic {\color{black} condition written in terms of its effective coupling constant satisfies ($\Omega /\omega_{0} \to \infty$), where $\Omega (\omega_{0})$ represents the atomic (mode) frequency.} Obviously, {\color{black} the variation of the effective coupling constant is determined by the atomic frequency and the mode frequency. The two frequencies are usually within a limited range in physical systems. Therefore, it is not straightforward to meet the thermodynamic condition $\Omega /\omega_{0} \to \infty$ readily in practical experiments.} {\color{black} There is no doubt that QPTs can result from diverse causes in different physical systems.
 In the systems, whether or not the critical point is approachable depends on the tunability of system parameters for the effective coupling constant to be as small as desired by the thermodynamic condition.} {\color{black} Therefore, with current experimental technologies, the QPT-assisted QSHE is still awaiting further explorations}, {\color{black} aiming to make it comparatively feasible in practical experiments by presenting different requirements according to the thermodynamic condition. One possible method towards the target is by expanding the physical model to revise the requirement(s), and moreover, the expansion can help to better understand the impact of QPT on the performance of QSHEs \cite{grimaudo2023thermodynamic}.}
	
 In this work, we {\color{black} utilize} two-qubit QRM with spin-spin coupling as a working medium of QSHE. Different from the commonly analyzed two-qubit QRM, {\color{black} our model includes} both qubit-mode interaction and qubit-qubit coupling, thereby enabling more complex interactions between the qubits. With spin-spin coupling, the conditions of QPT in this model change remarkably, {\color{black} and that} are determined by the infinite ratios of the spin-spin and the spin-mode couplings to the mode frequency, 
  {\color{black}  rather than the spin-to-mode frequency ratios \cite{grimaudo2023thermodynamic}.} This alteration affects the {\color{black} variation pattern} of the effective coupling constant toward the critical point, {\color{black}and this change may influence the overall performance of QSHE.} {\color{black} We then study the detailed effects on the performance of the QSHE} near the critical point by analyzing the intrinsic relationships between QPT and spin-spin coupling, spin-mode coupling strength, and mode frequency.  
{\color{black} To be specific,} our results indicate that in the normal phase, increasing the ratio of the hot and cold reservoir temperatures and the spin-mode coupling strength can improve the efficiency. {\color{black} Unfortunately, an increase in the spin-spin coupling strength does not play a positive role in improving} the efficiency. {\color{black} While} in the superradiant phase, the efficiency can approach the Carnot limit by lowering down the cold reservoir temperature or increasing the ratio of the spin-mode coupling strength to the mode frequency. 

 {\color{black} The paper is organized as follows.} {\color{black} In Sec. II, we explain the two-qubit QRM with spin-spin coupling and present the necessary thermodynamic requirements for observing QPT.} In Sec. III, we explore the properties of QSHE by using such a model as a working medium and analyzing the influence of QPT on its efficiency. In Sec. IV, we specifically investigate the effects of the coupling strength between spin-spin and spin-mode, as well as the frequency of the mode, on the efficiency of QSHE. {\color{black} In addition, we propose a feasible experimental scheme for realizing the two-qubit QRM with spin-spin coupling in a system of trapped ions in Sec.V. } Finally, we conclude in Sec. VI with a summary.

\section{The Two-qubit QRM with Spin-Spin Coupling}
	
In the two-qubit QRM with spin-spin coupling, two qubits interact not only with each other but also with a single bosonic field mode. The Hamiltonian for this model is given by the following equation ($\hbar=1$),
 \begin{equation}\label{eq1}
		\hat{H}=\omega _{0}\hat{a} ^{\dagger}\hat{a} +\varepsilon _{1} \hat{\sigma}_{1}^{z}+\varepsilon _{2} \hat{\sigma}_{2}^{z}+\gamma\hat{\sigma} _{1}^{x}\hat{\sigma}_{2}^{x}+(\lambda _{1}\hat{\sigma} _{1}^{z}+\lambda _{2}\hat{\sigma}_{2}^{z})(\hat{a}^{\dagger}+\hat{a}),
	\end{equation}
where $a^{\dagger}$ and $a$ are the annihilation and creation operators for a harmonic oscillator with frequency $\omega _{0}$. Here $\hat{\sigma}_{i}^{l}$ is the pauli matrix of the $i$-th qubit system $(i=1,2,l=x,y,z)$, and the transition frequency is $\varepsilon _{i}$. $\gamma$ and $\lambda _{i} (i=1,2)$ are the spin-spin and the $(i\mathrm{th})$ spin-mode couplings, respectively.

Under the canonical (symmetry) transformation \cite{PhysRevLett.107.100401}, the Hamiltonian remains invariant:
\begin{equation}\label{eq2}
\sigma _{i}^{x}\to -\sigma _{i}^{x},\sigma _{i}^{y}\to -\sigma _{i}^{y},\sigma _{i}^{z}\to \sigma _{i}^{z}, \quad i=1,2.
\end{equation}
This invariance {\color{black} shows for a unitary operator $\hat{\sigma} _{1}^{z}\hat{\sigma} _{2}^{z}$ corresponding to a $\pi$ rotation about the z-axis, two distinct subdynamics arise, each linked to the corresponding eigenvalues of the unitary operator $\hat{\sigma} _{1}^{z}\hat{\sigma} _{2}^{z}$.} Therefore, the total Hilbert space $\hat{H}$ can be described as the direct sum of two orthogonal subspaces \cite{PhysRevLett.130.043602},
$\hat{H}=\hat{H}_{+}\oplus \hat{H}_{-} $ with
	\begin{equation}\label{eq3}
		\hat{H}_{\pm } =\omega _{0}\hat{a} ^{\dagger}\hat{a} +\varepsilon _{\pm }\hat{\sigma}^{z}+\gamma\hat{\sigma}^{x}+\lambda _{\pm}(\hat{a}^{\dagger}+\hat{a})\hat{\sigma}^{z},
	\end{equation}
	where $\varepsilon _{\pm }=\varepsilon _{1}\pm \varepsilon _{2}$, $\lambda _{\pm}=\lambda _{1}\pm\lambda _{2}$, $\hat{\sigma}^{l}(l=x,y,z)$ is the standard two-level Pauli operator. The Hamiltonian $\hat{H}_{+}(\hat{H}_{-})$ of the two-qubit QRM with spin-spin coupling is projected onto the orthogonal subspace $\mathcal{H}_{+}(\mathcal{H}_{-})$, whose basis consists of a specific quantum state and a mode Fock basis. 
	{\color{black} Virtual state mapping reduce the Hamiltonian $\eqref{eq3}$ to two independent QRMs, with the {\color{black} qubit-qubit} coupling $\gamma$ simulating transverse fields. This setup, unlike $\eqref{eq1}$, can be easily achieved with external classical control. {\color{black} In this work, we examine counter-biased qubits, where $\varepsilon_1 = -\varepsilon_2 = \frac{\varepsilon}{2}$ and $\lambda_1 =\lambda_2 = \frac{\lambda}{2}$.} Thus, Hamiltonian $\eqref{eq3}$ can be rewritten as,
	\begin{equation}\label{eq4}
		\begin{aligned} 
			\hat{H}_{+} &= \omega _{0}\hat{a} ^{\dagger}\hat{a} +\gamma\hat{\sigma}^{x}+\lambda(\hat{a}^{\dagger}+\hat{a})\hat{\sigma}^{z},  \\ 
			\hat{H}_{-} &=\omega _{0}\hat{a} ^{\dagger}\hat{a} +\varepsilon\hat{\sigma}^{z}+\gamma\hat{\sigma}^{x}.
		\end{aligned}
	\end{equation}
	{\color{black} The Hamiltonian $\hat{H}_{+}$ corresponds to the "standard" single-spin QRM, while $\hat{H}_{-}$ describes a two-level system decoupled from the boson field. After rotating the system by $\frac{\pi}{2}$ around the y-axis, the two Hamiltonians in $\eqref{eq4}$ become,}
	\begin{equation}\label{eq5}
		\begin{aligned} 
			\hat{H}_{+} &= \omega _{0}\hat{a} ^{\dagger}\hat{a} +\gamma\hat{\sigma}^{z}+\lambda(\hat{a}^{\dagger}+\hat{a})\hat{\sigma}^{x}, \\ 
			\hat{H}_{-} &=\omega _{0}\hat{a} ^{\dagger}\hat{a} +\varepsilon\hat{\sigma}^{x}+\gamma\hat{\sigma}^{z}.
		\end{aligned}
	\end{equation}
    {\color{black} As known in \cite{grimaudo2023thermodynamic}, this two-qubit QRM with spin-spin coupling exhibits a superradiant quantum phase transition (QPT) in the thermodynamic limit with ($\gamma /\omega_{0} \to \infty$, $\lambda /\omega_{0} \to \infty$). Compared to the traditional QRM,  this phase transition is driven by the infinite ratios of spin-spin and spin-mode couplings to the mode frequency, rather than by the spin-to-mode frequency ratio. }
To determine the critical point, it is necessary to obtain the effective Hamiltonian of  $\eqref{eq5}$ by Schrieffer-Wolf transformation \cite{PhysRevLett.115.180404}, which can be given by $H_{\mathrm{eff}}=\hat{U}^{\dagger}\hat{H}_{\pm}\hat{U}$ with $\hat{U}=\exp [{\frac{\lambda }{2\gamma}}(a+a^{\dagger})i\sigma^{y}]$.
 Projecting the transformed Hamiltonian onto the spin-down subspace yields the minimum effective Hamiltonian.  
 In the normal phase, the exact low-energy form of the
Hamiltonian is
	\begin{equation}\label{eq6}
		\hat{H}_{+}^{np} =\omega_{0} a^{\dagger}a-\frac{\omega_{0}}{4}g^{2}(a^{\dagger}+a)^{2}-\gamma.
	\end{equation}
{\color{black} Through diagonalization, the Hamiltonian is also written as }$\hat{H}_{+}^{np} = \varepsilon _{np}(g)-\gamma$ with the corresponding energy $\varepsilon _{np}(g) =\omega_{0}\sqrt{1-g^{2}}$, where $\varepsilon _{np}(g)$ is real only for $g< 1$. 
In the superradiant phase, where $g> 1$, the low-energy effective Hamiltonian reads
	\begin{equation}\label{eq7}
		\hat{H}_{+}^{sp} =\omega_{0} a^{\dagger}a-\frac{\omega_{0}}{4g^{4}}(a^{\dagger}+a)^{2}-\gamma\frac{g^{2}+g^{-2}}{2}. 
	\end{equation}
 {\color{black} Upon diagonalizing, the Hamiltonian can be written as $\hat{H}_{+}^{sp} =\varepsilon _{sp}(g)-\gamma\frac{g^{2}+g^{-2}}{2}$ with the corresponding eigenenergy} $\varepsilon_{sp}(g)=\omega_{0}\sqrt{1-g^{-4}}$, where $g$ serves as an effective coupling constant. This effective coupling constant $g$, is defined as \( g = \frac{\sqrt{2}\lambda}{\sqrt{\omega_{0} \gamma}} =\frac{\sqrt{2}\xi}{\sqrt{\zeta}} \), with $\xi$ and $\zeta$  being dimensionless ratios defined as $\xi=\frac{\lambda}{\omega_{0}}$ and $\zeta=\frac{\gamma}{\omega_{0}}$, respectively.
{\color{black} The imaginary single-qubit QRM} Hamiltonian $\hat{H}_{+}$ in $\eqref{eq5}$ exhibits a second-order QPT from the normal phase to the superradiant phase, driven by the effective coupling constant 
$g$. This behavior implies that the two-qubit QRM with spin-spin coupling undergoes a QPT under certain conditions.
	\section{THE QSHE}
	\subsection{The Stirling Cycle with Two-Qubit QRM }
 We first discuss a QHE model based on the Stirling cycle, which utilizes a two-qubit QRM with spin-spin coupling as a working medium. The cycle, depicted in Fig. 1, consists of two quantum isothermal processes (A-B and C-D) and two isochoric processes (B-C and D-A).
 \begin{figure}[htp]
		\centering
		\includegraphics[width=1\columnwidth]{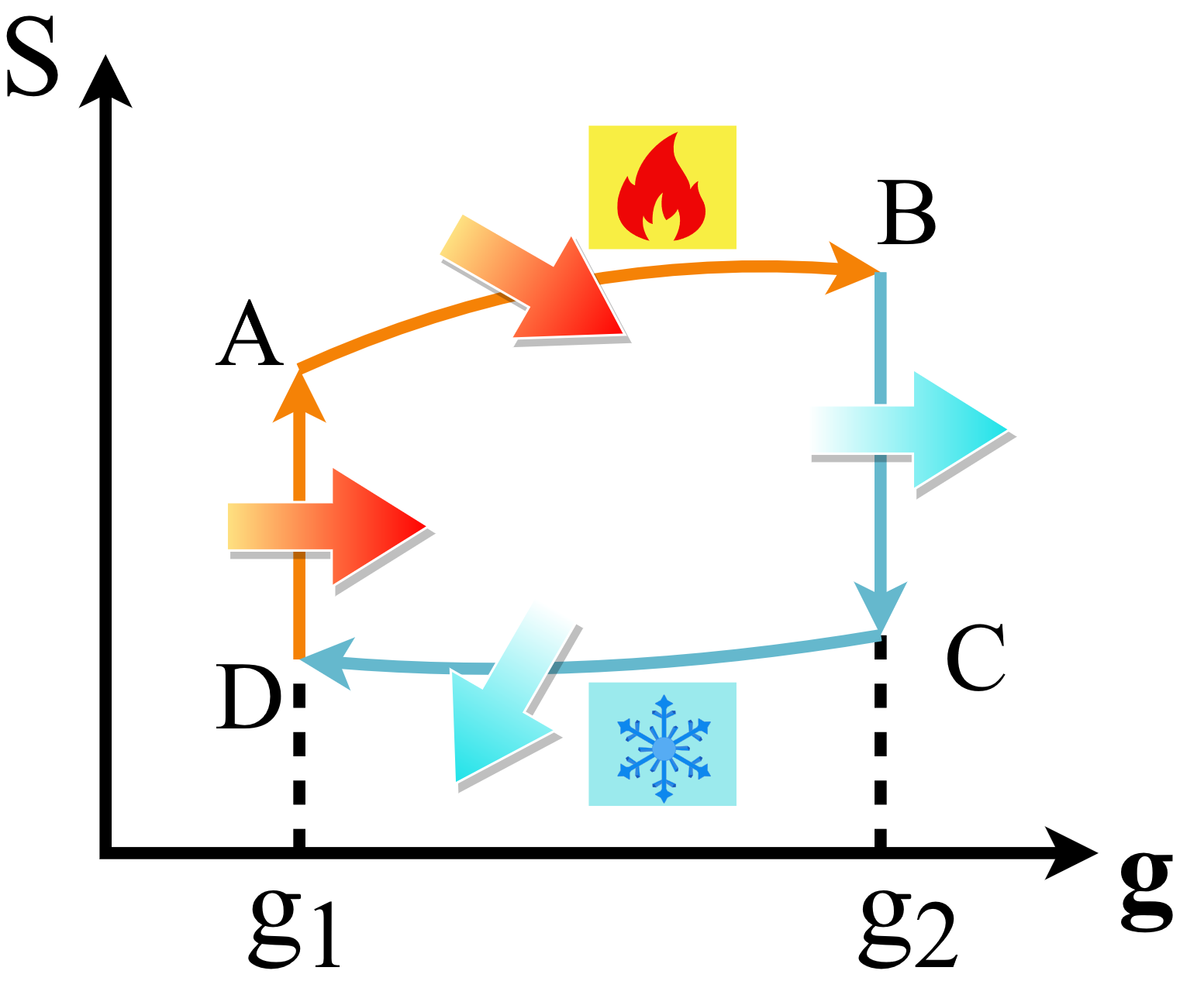}
		\caption{Diagram of the entropy-coupling for a two-qubit quantum Rabi-Stirling heat engine cycle. In this diagram, \( g_{1} \) and \( g_{2} \) represent the effective coupling constants during the isochoric processes from \( D \) to \( A \) and from \( B \) to \( A \), respectively. \( T_{H} \) and \( T_{C} \) denote the temperatures of the hot and cold reservoirs, respectively.}
		\label{wildlife}
	\end{figure}
	
    (1). The isothermal compression process (A-B): In the first stroke, the effective coupling constant, $g$, slowly increases from $g_{1}$ to $g_{2}$. During this process, the system {\color{black} is connected to} a hot reservoir at temperature $T_{H}$. {\color{black} With the quasi-static assumption that the energy of the system changes much more slowly than the relaxation rates do, we suppose }that the system remains in thermal equilibrium with the reservoir. {\color{black} As a result}, the heat gained} by the system from the hot reservoir can be expressed as $Q_{AB}=T_{H}\bigtriangleup S_{AB}$, where $\bigtriangleup S_{AB}$ denotes the change in entropy of the system.

	(2). Isochoric compression process (B-C): In the second stroke, the system is disconnected from the hot reservoir and {\color{black} rapidly comes in contact with the cold reservoir at $T_{C}$, while managing to maintain} $g=g_{2}$. During this process, no work {\color{black} is done}. Consequently, the heat transferred between the system and the cold reservoir equals the change in the system's internal energy, $Q_{BC}=U_{C}-U_{B}$.
 
	(3). The isothermal expansion process (C-D): In the third stroke, the system remains  {\color{black} connected to} the cold reservoir while the variable $g$ is slowly decreased from $g_{2}$ to $g_{1}$. During this process, the system  {\color{black} dissipates} heat to the cold reservoir. The amount of heat {\color{black} dissipates} is given by  $Q_{CD}=T_{C} \bigtriangleup S_{CD}$, where $\bigtriangleup S_{CD}$ is the entropy change of the system.
 
	(4). The isometric expansion process (D-A): {\color{black} At the final stage of the cycle, the system loses touch with the cold reservoir and comes in contact with a hot reservoir at temperature $T_{H}$, while managing to keep the effective coupling constant $g_{1}$ unchanged. During this process, nothing does work.} The heat exchanged between the system and the hot reservoir is equal to the change in the system's internal energy, $Q_{DA} =U_{A}-U_{D}$.

{\color{black} Assume} that all thermodynamic processes are quasistatic, the equilibrium state of the system is described by the Gibbs state $\rho = \frac{1}{\mathcal{Z}} \exp\left[-\frac{H_{+}^{np}}{T}\right]$,
where the partition function is given by $ \mathcal{Z} = \text{Tr}\left[\exp\left(-\frac{H_{+}^{np}}{T}\right)\right]$, and $T$ is the temperature (with the Boltzmann constant $k_B = 1$). The internal energy and entropy of the system in equilibrium are $ U = \frac{\varepsilon_{np}(g)}{\exp(\varepsilon_{np}(g)/T) - 1} + E_0$ and $S = \frac{\varepsilon_{np}(g)/T}{\exp(\varepsilon_{np}(g)/T) - 1} - \ln{\left[1 - \exp\left(-\frac{\varepsilon_{np}(g)}{T}\right)\right]}$ respectively.

Heat absorption occurs during the first isothermal stroke and the final isochoric stroke, so the total heat {\color{black} gained} is \( Q_{\text{in}} = Q_{AB} + Q_{CD} \). The total work done is \( W = Q_{AB} + Q_{BC} + Q_{CD} + Q_{DA} \). Hence, the cycle efficiency \( \eta = \frac{W}{Q_{\text{in}}} \) can be expressed as,
\begin{equation}\label{eq8}
    \eta = \frac{\eta_c + \Sigma_1 + \Sigma_2}{1 + \Sigma_2},
\end{equation}
where \( \eta_c = 1 - \frac{T_C}{T_H} \) is the Carnot efficiency, and the terms 
$\Sigma_1 = \frac{Q_{BC} - T_C(\Delta S_{DA} + \Delta S_{BC})}{Q_{AB}}$ and $\Sigma_2 = \frac{Q_{DA}}{Q_{AB}}$ account for the irreversibilities in the Stirling cycle.

The Carnot cycle, consisting of two isothermal and two adiabatic processes, defines the maximum theoretical efficiency. {\color{black} While the Stirling cycle in which the adiabatic processes are replaced by isochoric ones, has an efficiency less than Carnot's due to these irreversible terms \( \Sigma_1 \) and \( \Sigma_2 \). To approach the Carnot efficiency, the irreversible effects in the Stirling cycle need to be minimized.}

\begin{figure*}[htpb]
        \centering
		\includegraphics[width=0.68\columnwidth]{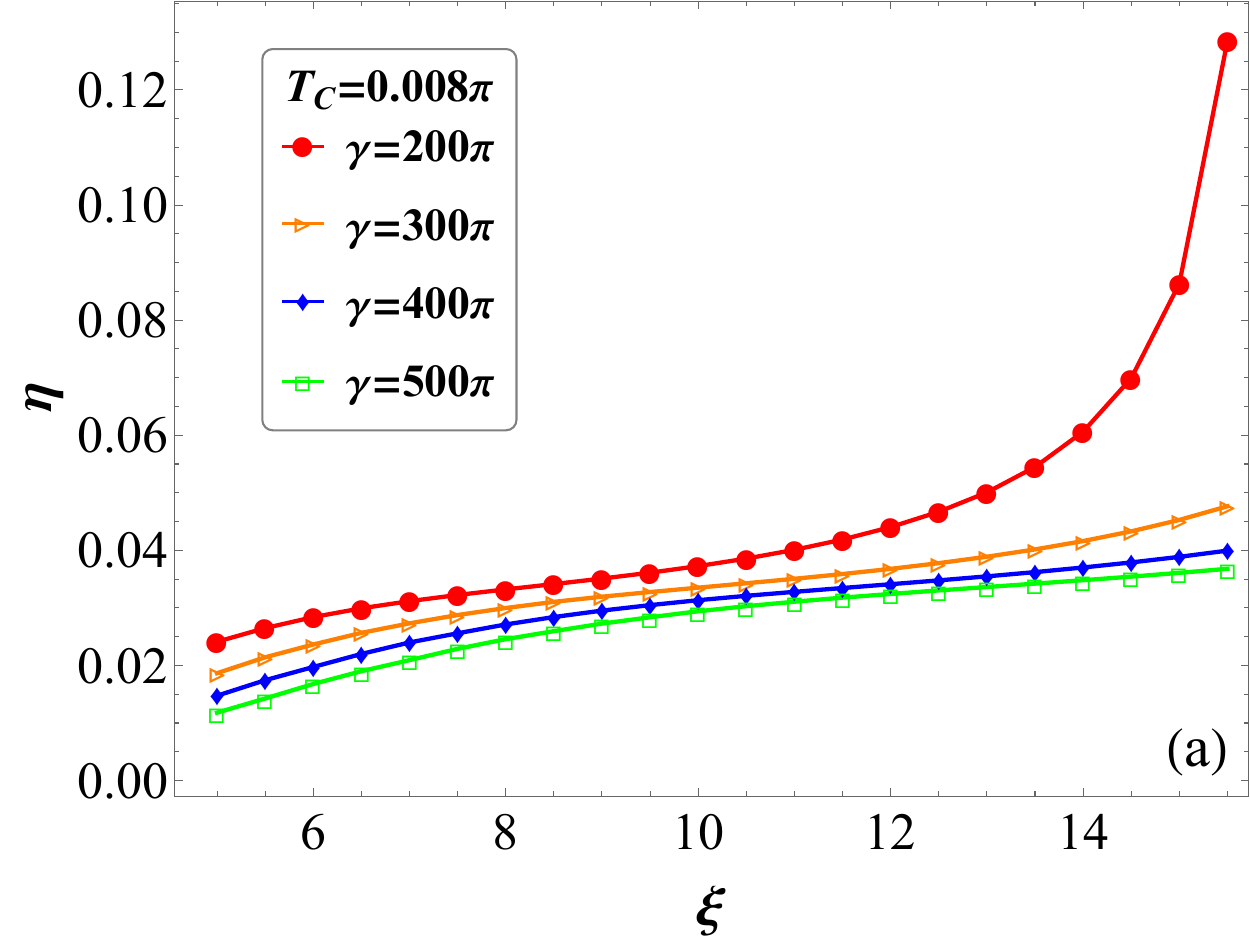}
		\includegraphics[width=0.68\columnwidth]{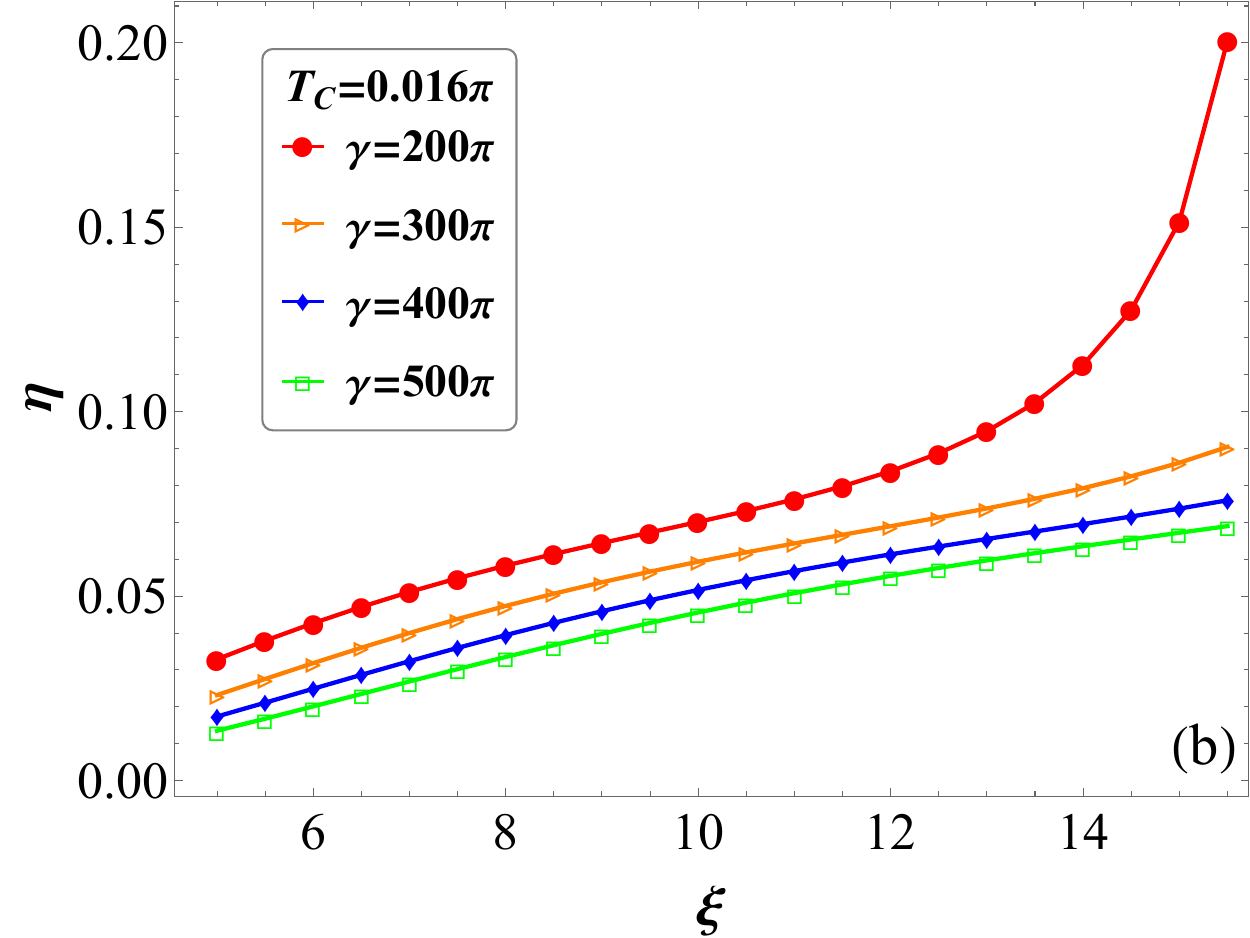}
		\includegraphics[width=0.68\columnwidth]{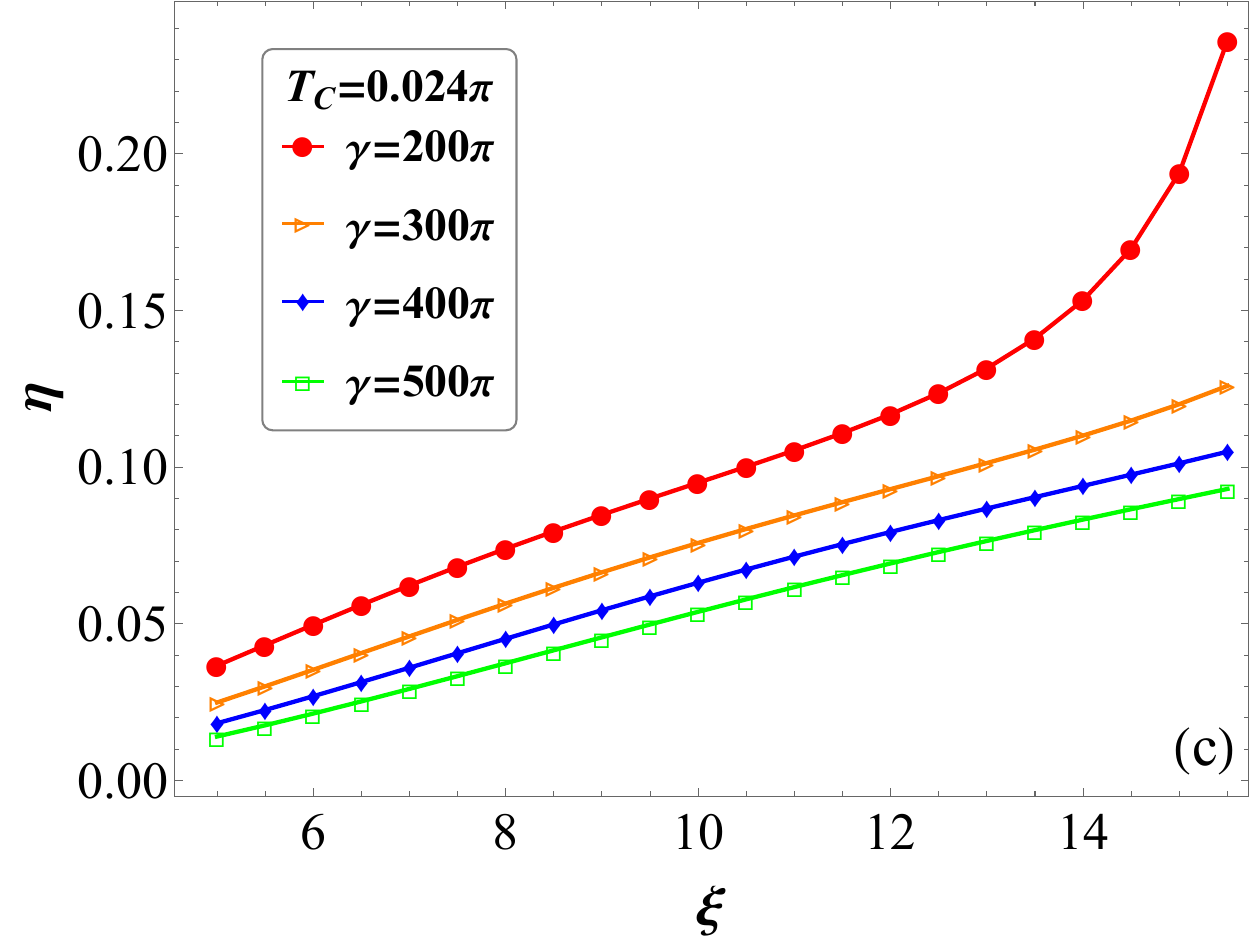}
        \includegraphics[width=0.68\columnwidth]{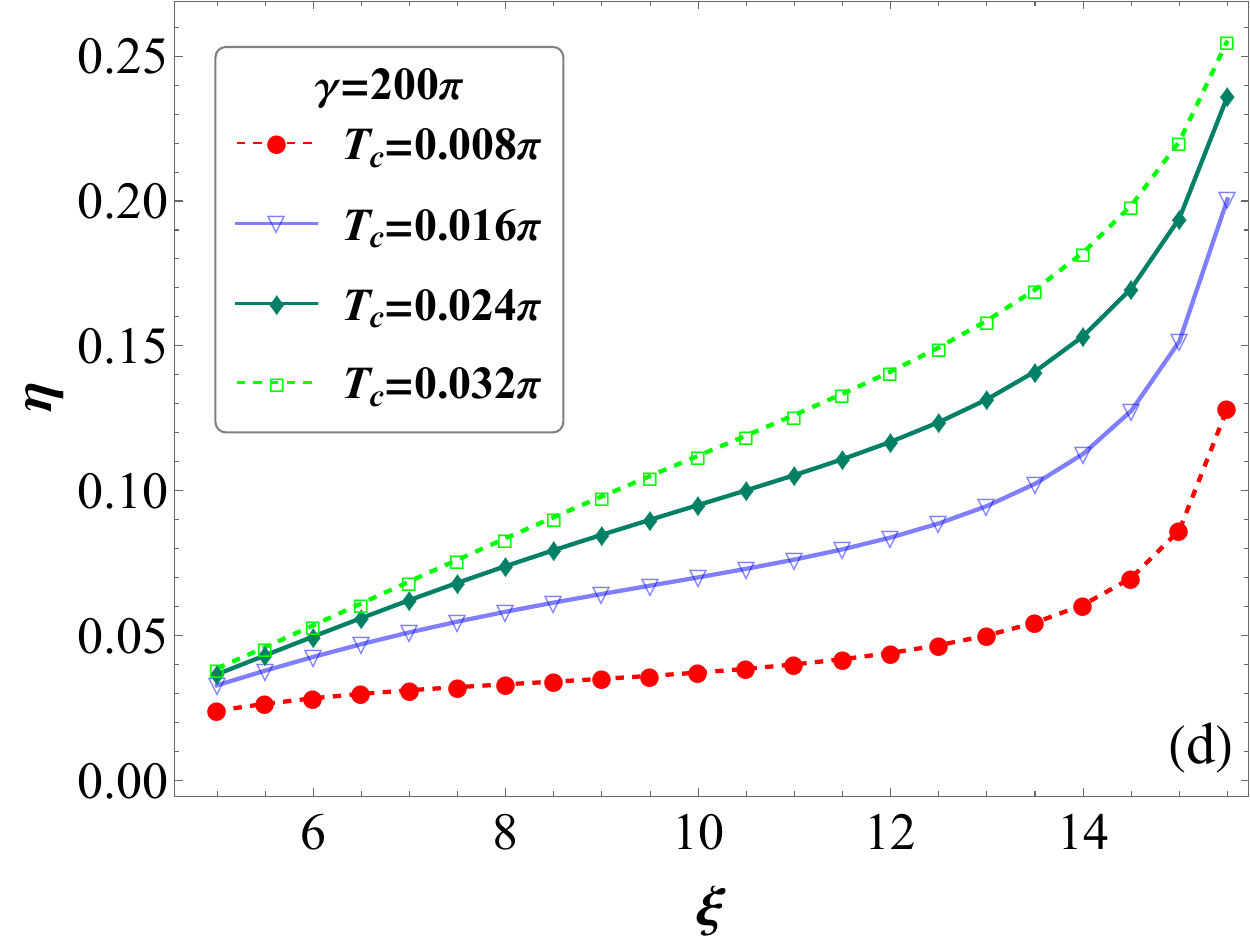}
	  \includegraphics[width=0.68\columnwidth]{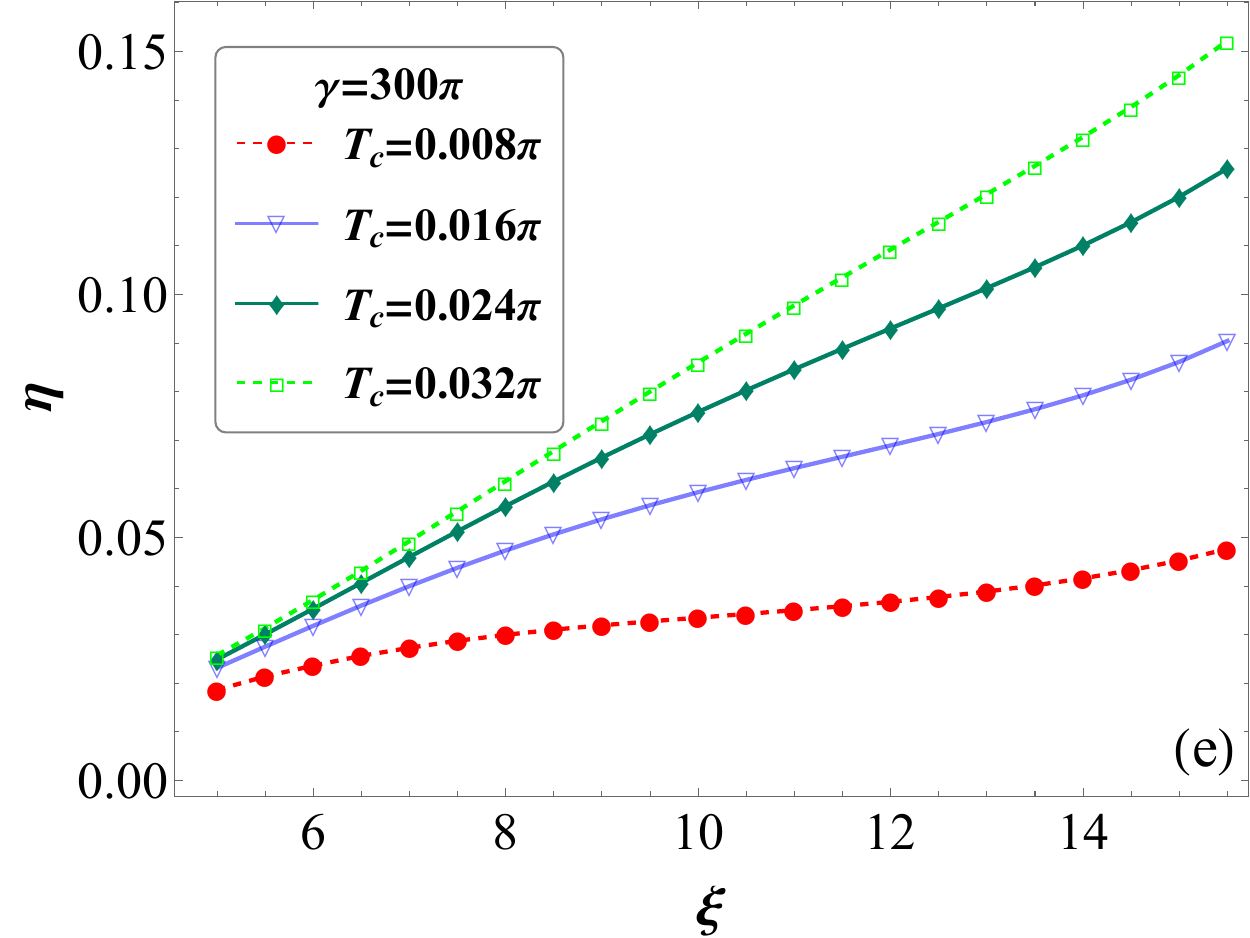}
	  \includegraphics[width=0.68\columnwidth]{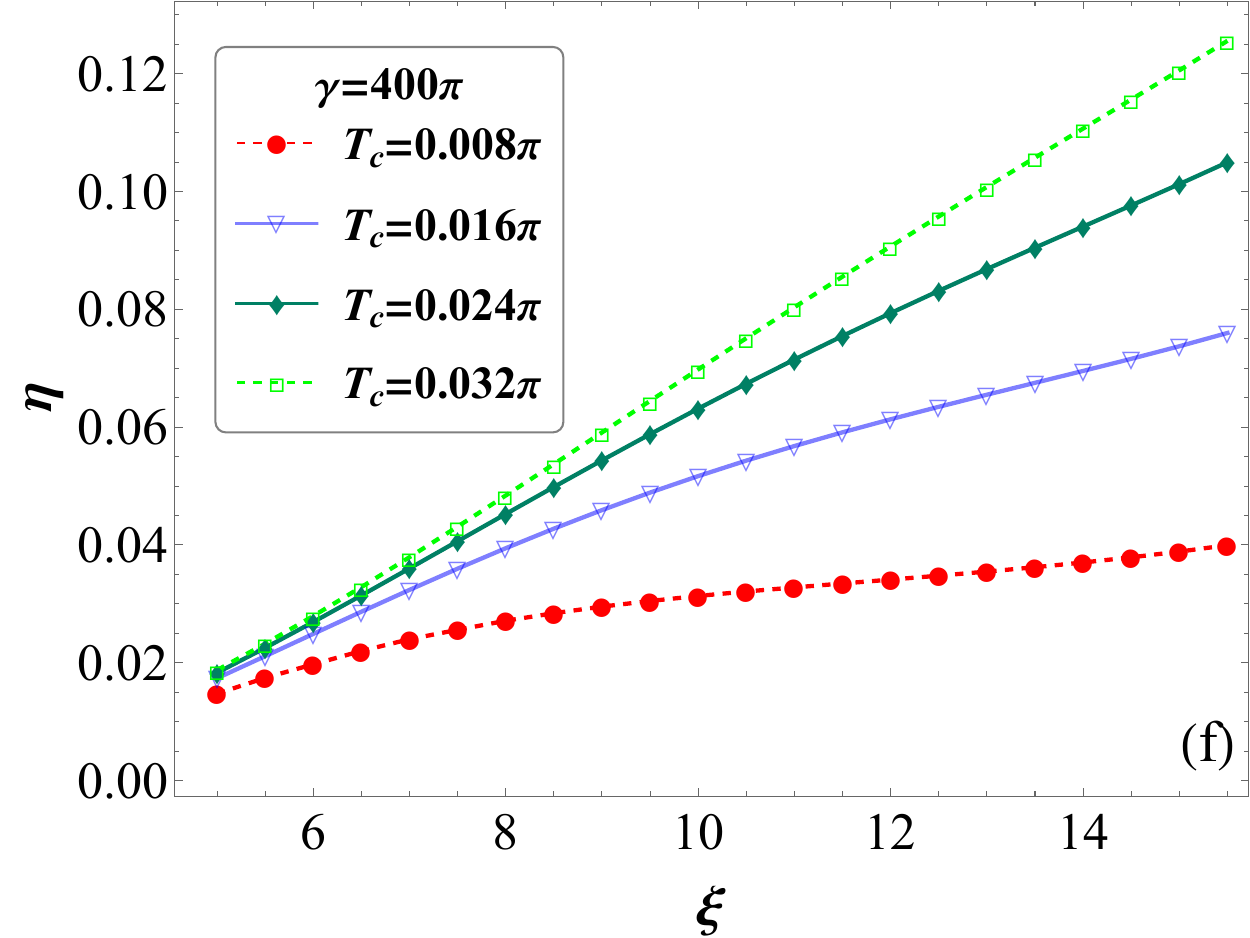}
		\caption{In the normal phase, the QSHE efficiency $\eta$ as a function of $\xi$.In (a)-(c),we show the variation of $\eta$ at different coupling strengths $\gamma=200 \pi$ (red solid line with solid circles), $\gamma=300 \pi$ (orange solid line with hollow triangles), $\gamma=400 \pi$ (blue solid line with solid diamonds), $\gamma=500 \pi$ (green solid line with hollow squares).The temperature is chosen as $T_{c}=0.008 \pi$ for (a), $T_{c}=0.016 \pi$ for (b) and $T_{c}=0.024 \pi$ for (c).In (d)-(f), we demonstrate the changing pattern of $\eta$ at different temperatures $T_{c}=0.008 \pi$ (red dotted line with solid circles), $T_{c}=0.016 \pi$ (the light blue solid line with hollow inverted triangles), $T_{c}=0.024 \pi$(the dark green solid line with solid diamonds), $T_{c}=0.032 \pi$ (green dotted line with hollow squares).In this case, we chose coupling strength $\gamma=200 \pi$ for (d), $\gamma=300 \pi$ for (e) and $\gamma=400 \pi$ for (f). The value of the parameter $\alpha$ is fixed to be $1.5$.}
		\label{wildlife}
	\end{figure*}	

\subsection{ The Efficiency of QSHE via Criticality}
 
{\color{black} In the following, we present a QSHE based on a two-qubit QRM with spin-spin coupling,} operating in the normal phase where the effective coupling constant \( g < 1 \). {\color{black} In this setting,} the effective coupling constant \( g \) is controlled by the spin-spin coupling strength \( \lambda \), spin-mode coupling strength \( \gamma \), or by the parameters \( \zeta \) and \( \xi \). Thus, \( \lambda \), \( \gamma \), \( \zeta \), and \( \xi \) are chosen as the key adjustable parameters in the thermodynamic cycle. {\color{black} We then explore the thermodynamic cycle and highlight in which conditions the efficiency approaches the Carnot limit, through adjusting the parameters to move the system closer to the critical point. Our method is similar to the one discussed in \cite{PhysRevA.109.022208}.}

{\color{black} To achieve the Stirling efficiency (denoted in $\eta$) as close to the Carnot limit (denoted in $\eta_c$) as desired,} the terms \( \Sigma_1 \) and \( \Sigma_2 \) in Eq. \eqref{eq8} should approach zero at the critical point. {\color{black} This requires that the numerators \( Q_{BC} \), \( \Delta S_{DA} \), \( \Delta S_{BC} \), and \( Q_{DA} \) remain finite, while the denominator \( Q_{AB} \) diverges as the system goes to }the critical point.

In the quantum isochoric process, {\color{black} where no work is done,} the heat exchanged with the reservoir equals the change in the system's internal energy. Focusing on heat absorption, the absorbed heat \( Q_{DA} \) can be expressed as,
\begin{equation}\label{eq9}
    Q_{DA} = \int_{T_H}^{T_C} \frac{\partial U(T, g_1)}{\partial T} \, dT \equiv \int_{T_H}^{T_C} \mathcal{C}\left[\frac{\varepsilon_{np}(g_1)}{T}\right] dT,
\end{equation}
where \( \mathcal{C}\left[\frac{\varepsilon_{np}(g_1)}{T}\right] \) is the isothermal heat capacity, 
 {\color{black} and \( \mathcal{C}(x) \) is given by $
\mathcal{C}(x) = \left(\frac{x}{2}\right)^2 \operatorname{csch}^2\left(\frac{x}{2}\right)
$. The \( \mathcal{C}(x) \) is a monotonically decreasing function over \( x \in (0, \infty) \), with limits \( \lim_{x \to 0} \mathcal{C}(x) \to 1 \) and \( \lim_{x \to \infty} \mathcal{C}(x) \to 0 \).} Therefore, \( Q_{DA} \) is finite and satisfies:
\begin{equation}\label{eq10}
    0 < Q_{DA} < (T_H - T_C) \mathcal{C}\left[\frac{\varepsilon_{np}(g_1)}{T_H}\right].
\end{equation}

In the quantum isochoric process, the derivative of entropy with respect to temperature is given by \( \frac{\partial S(T, g)}{\partial T} = \frac{\mathcal{C}[\varepsilon_{np}(g)/T]}{T} \). Thus, the change in entropy \( \Delta S_{DA} \) can be expressed as,
\begin{equation}\label{eq11}
    \Delta S_{DA} = \int_{T_C}^{T_H} \frac{1}{T} \mathcal{C}\left[\frac{\varepsilon_{np}(g_1)}{T}\right] dT,
\end{equation}
and $0 < \Delta S_{DA} < \frac{T_H - T_C}{T_C} \mathcal{C}\left[\frac{\varepsilon_{np}(g_1)}{T_H}\right]$.

Similarly, it can be shown that \( Q_{BC} \) and \( \Delta S_{BC} \) are also finite in the quantum isochoric process, $-(T_H - T_C) \mathcal{C}\left[\frac{\varepsilon_{np}(g_2)}{T_H}\right] < Q_{BC} < 0$ and $  -\frac{T_H - T_C}{T_C} \mathcal{C}\left[\frac{\varepsilon_{np}(g_2)}{T_H}\right] < \Delta S_{BC} < 0$.

{\color{black} From the above-discussed results,} it is confirmed that \( Q_{BC} \), \( \Delta S_{DA} \), \( \Delta S_{BC} \), and \( Q_{DA} \) are bounded {\color{black} by finite values}. Next, we investigate \( Q_{AB} \) as \( g_2 \) approaches the critical point. As the energy gap closes, \( \varepsilon_{np}(g_2) \to 0 \), and the entropy \( S_B \) can be approximated as,
\begin{equation}\label{eq12}
    S_B \approx -\ln\left[\frac{\varepsilon_{np}(g_2)}{T_H}\right].
\end{equation}
Thus, \( S_B \) and \( S_C \) diverge near the critical point, making \( \Delta S_{BC} \) finite. In the quantum isochoric process \( D \to A \), \( S_A \) remains finite, so, 
\begin{equation}\label{eq13}
    \lim_{g_2 \to 1} Q_{AB} \equiv T_H (S_B - S_A) \to \infty.
\end{equation}
Finally, it follows that,
\begin{equation}\label{eq14}
    \lim_{g_2 \to 1} \Sigma_1 = \lim_{g_2 \to 1} \Sigma_2 \to 0.
\end{equation}

{\color{black} The result shown in Eq. (14) demonstrates} that when \( g_1 < g_2 \to 1 \), the efficiency $\eta$ is {\color{black} approaching the Carnot limit $\eta_c$.} It is important to note that this result is only applicable at finite temperatures. The low-energy effective Hamiltonian used here, based on the two-qubit QRM with spin-spin coupling, may not accurately capture the system's dynamics at high energy levels or in the high-temperature limit, where additional factors become significant. Therefore, this study does not explore the system's behavior in these regimes \cite{PhysRevA.109.022208}.

 {\color{black} We have shown that as long as $Q_{AB}$ diverges, the efficiency $\eta$ goes to the Carnot limit $\eta_c$. Some more details are provided as follows.} First, we make the ratio of the hot and cold reservoir temperatures $\alpha=T_{H}/T_{C}$ a finite value. When the system is at the low-temperature limit ( $T_{H}\to 0 $ ), the system is held in the ground state. In the normal phase, the ground state of the system is non-degenerate, so $\textstyle \lim_{T_{H} \to 0}S_{A}\to 0$ and $\textstyle \lim_{T_{H} \to 0}S_{D}\to 0$. In the superradiant phase, the ground state is in double degeneracy, so $\textstyle \lim_{T_{H} \to 0}\Delta S_{AB}\to S_{B}=\ln{2} $ in the low-temperature limit. Similarly, in the superradiant phase, we can still derive the $\Delta S_{BC}$ and $Q_{BC}$ limits, $-(T_{H}-T_{C})\mathcal{C}[\varepsilon _{sp}(g_{2})/T_{H}]< Q_{BC} <0$ and $-\frac{T_{H}-T_{C}}{T_{C}}\mathcal{C}[\varepsilon _{sp}(g_{2})/T_{H}]<\Delta S_{BC} < 0$. In this context, we find that $\lim_{T_{H} \to 0} \eta \to \eta_{c}$, indicating that in the low-temperature limit, the efficiency for $g_{1} < g_{2} \to 1$ approaches the Carnot limit.

Next, we examine how the efficiency $\eta$  approaches the Carnot limit $\eta_{c}$ as the effective coupling constant $g_{2}$ goes to the critical point. As shown in Eq. \eqref{eq13}, when \(g_{2}\) goes closer to the critical point, both \(\textstyle \sum_{1}\) and \(\textstyle \sum_{2}\) tend to zero, allowing Eq. \eqref{eq14} to be approximated by $\eta - \eta_{c} \approx \frac{\Lambda}{\Delta S_{AB}}$
{\color{black} where \(\Lambda \equiv (1 - \eta_{c}) \left(\frac{Q_{DA}}{T_{H}} + \frac{Q_{BC}}{T_{C}} - \Delta S_{DA} - \Delta S_{BC}\right)\). Obviously, \(\Lambda\) is bounded by finite values. Therefore,}
\begin{equation}\label{eq15}
    \lim_{g_{2} \to 1} \eta \to \eta_{c} - \frac{\Lambda}{\ln{\left[\frac{\varepsilon_{np}(g_{2})}{T_{H}}\right]}}.
\end{equation}
The energy gap \(\Delta\) at the critical point follows a universal law \(\Delta \propto |g - g_{c}|^{zv}\), with \(zv = 1/2\), where $z$ and $v$ are the critical exponents of the QPT. Near the critical point, \(\varepsilon_{np}(g_{2}) \propto |g_{2} - g_{c}|^{1/2}\). Thus, Eq.\eqref{eq15} can be rewritten as
\begin{equation}\label{eq16}
    \lim_{\frac{\sqrt{2}\lambda_{2}}{\sqrt{\omega_{0} \gamma_{2}}} \to 1} \eta \to \eta_{c} - \frac{\Lambda}{\ln{\left| \frac{\sqrt{2}\lambda_{2}}{\sqrt{\omega_{0} \gamma_{2}}} - 1 \right|^{1/2}}}.
\end{equation}
{\color{black} As \(g_{2}\) is closer and closer to 1, the efficiency \(\eta\) is moving toward the Carnot limit \(\eta_{c}\).} Various strategies can be employed to bring \(g_{2}\) closer to 1 and thereby improve the heat engine's efficiency. {\color{black} We then conduct a quantitative analysis to evaluate how these methods impact efficiency.}

\begin{figure*}[htpb]
 \centering
	\includegraphics[width=0.72\columnwidth]{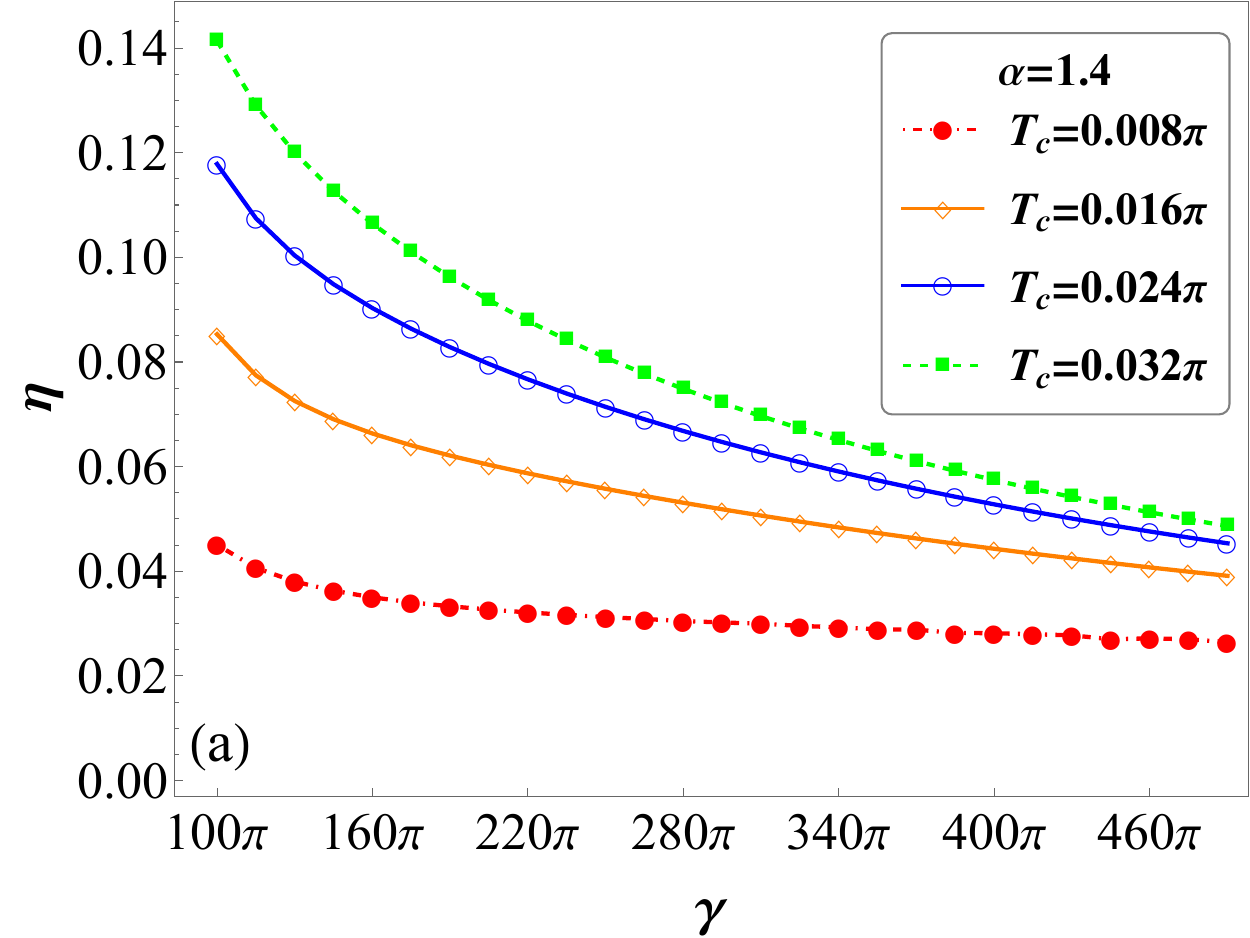}
	\includegraphics[width=0.72\columnwidth]{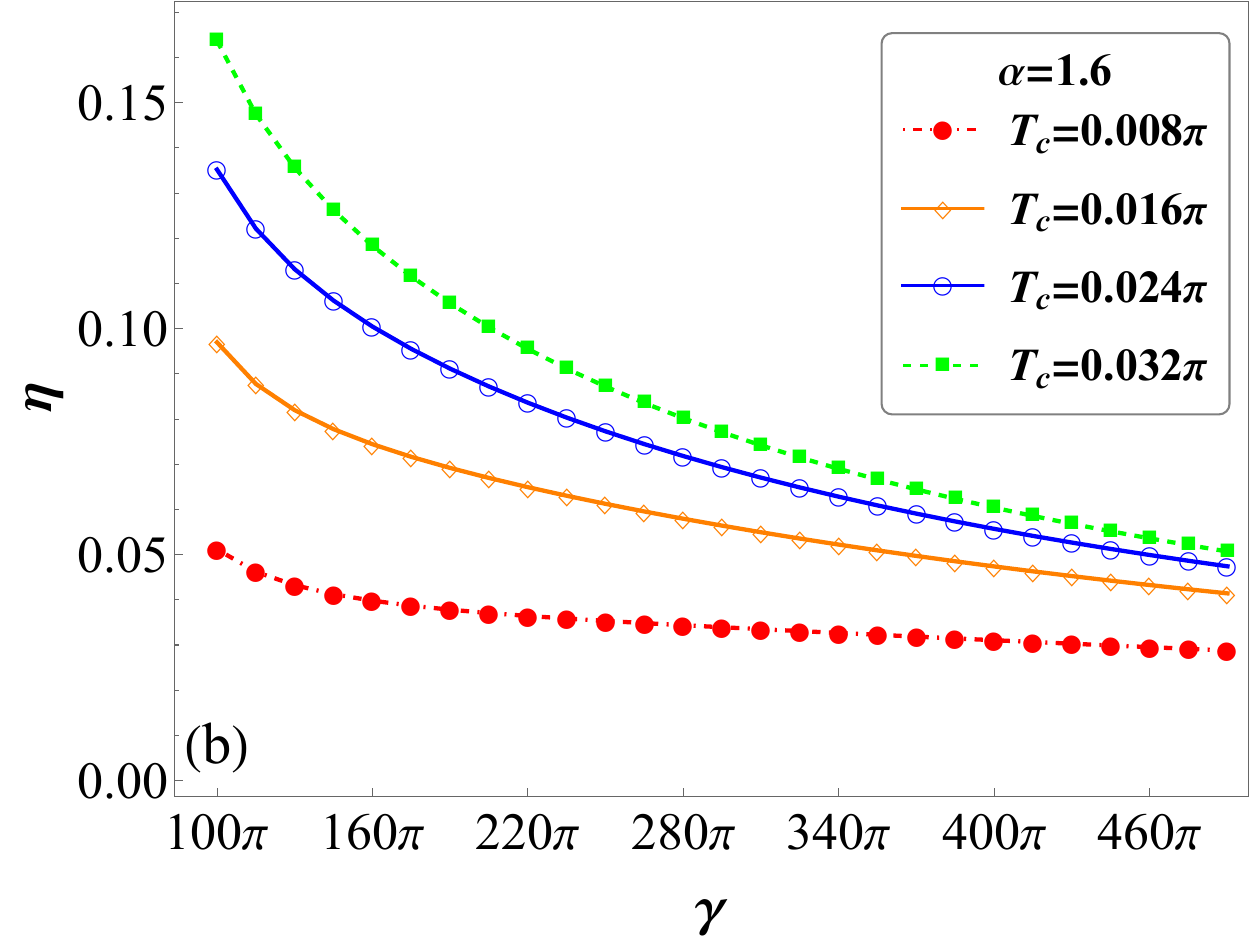}
	\includegraphics[width=0.72\columnwidth]{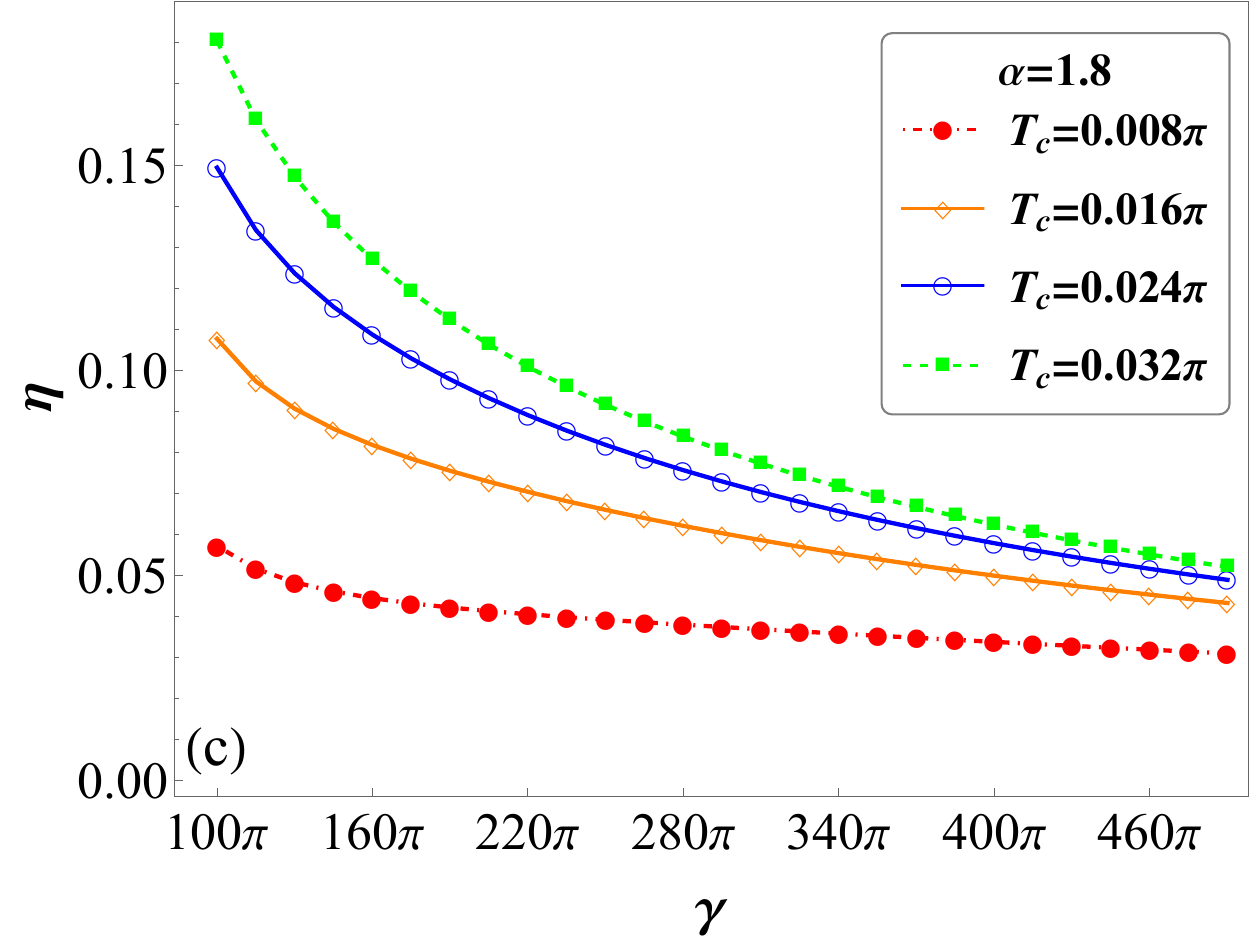}
    \includegraphics[width=0.72\columnwidth]{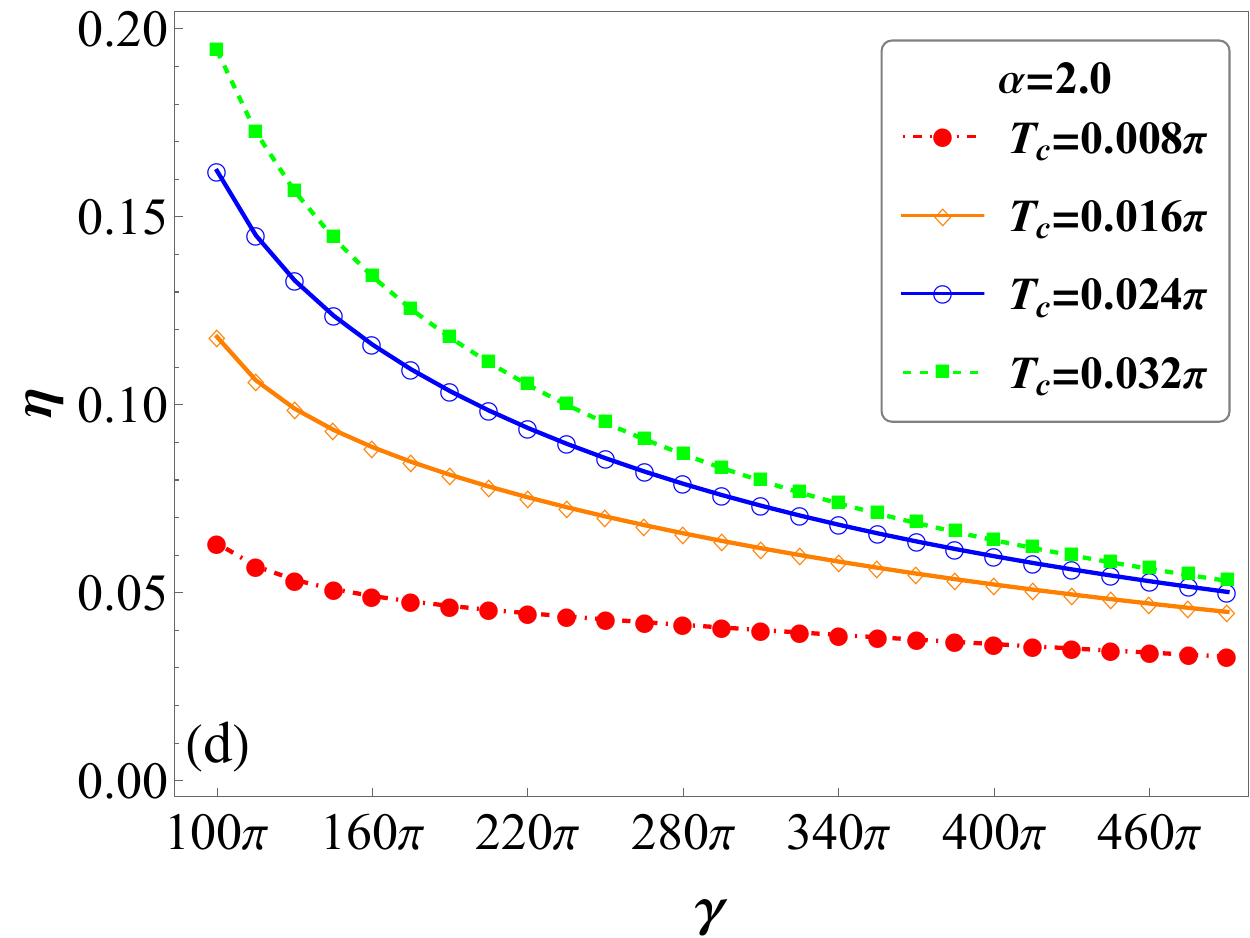}
	\caption{In the normal phase, the efficiency of the QSHE $\eta$ as a function of the strength of spin-spin coupling $\gamma$ at different temperatures $T_{c}=0.008 \pi$ (red dotted line with solid circles), $T_{c}=0.016 \pi$ (orange dotted line with hollow diamonds), $T_{c}=0.024 \pi$ (blue dotted line with hollow circles), and $T_{c}=0.032 \pi$ (green dotted line with solid squares). The temperature ratio of hot-to-cold reservoirs $\alpha$ = 1.4 for (a), $\alpha$ = 1.6 for (b), $\alpha$ = 1.8 for (c), and $\alpha$ = 2.0 for (d). Moreover, we select the parameter $\xi= 9$.}
	\label{wildlife}
   \end{figure*} 
 
	\section{Numerical Analysis of the Performance of the QSHE}
 
{\color{black} According to Ref. \cite{grimaudo2023thermodynamic}, QPT conditions in a two-qubit QRM with spin-spin coupling are primarily determined by two key ratio values, one is the spin-spin coupling strength to mode frequency and the other one is the spin-mode coupling strength to mode frequency. The critical point can be approached along different trajectories by adjusting the values. As a result, the performance of the QSHE is apparently affected by these ratio values and the system parameters. Moreover, the effective coupling constant $g$ also controls the conditions for observing QPT. Thus, it exerts influence on the performance of the QSHE too, especially for the energy conversion efficiency and the stability of QSHE. }


{\color{black} We numerically analyze the properties of the QSEH when the system is moving toward its critical point via different paths tuned by changing the system parameters. We know from the previous discussions that} the total heat absorbed during the cycle is $Q_{in} = Q_{AB} + Q_{CD}$, and the total work done is $W = Q_{AB} + Q_{BC} + Q_{CD} + Q_{DA}$, leading to an efficiency $\eta = W/Q_{in}$. In our calculations, we choose $\omega_{0} = 0.4 \pi  \mathrm{kHz}$, $g_{1} = 0.1$, and $\frac{\sqrt{2}\lambda}{\sqrt{\omega_{0} \gamma }}\ge 0.1$, and $\alpha = T_{H}/T_{C}$. The parameters used here are from recent theoretical and experimental studies on the QPTs in the QRM \cite{PhysRevLett.118.073001}.

{\color{black}The impact of three system parameters} on the efficiency of the QSHE is thoroughly analyzed. Fig. 2, we illustrate how the efficiency ($\eta$) of the QSHE varies with the parameter $\xi$ across different coupling strengths in the normal phase. The analysis reveals that the efficiency grows significantly as the ratio of spin-mode coupling strength to mode frequency ($\xi$) increases. However, {\color{black}a larger and larger} spin-spin coupling strength ($\gamma$) gradually reduces efficiency, as shown in Figs. 2(a)–2(c). Similarly, the temperature of the cold reservoir ($T_C$) plays a critical role {\color{black} and that is as $T_C$ increases,} the efficiency rises. Figs. 2(d)–2(f) show that with increasing $\gamma$, the QSHE efficiency decreases and becomes more linear. Hence, within the normal phase, the spin-spin coupling strength reduces the efficiency of the QSHE, while both the temperature and the ratio of spin-mode coupling strength to mode frequency enhance it.

   {\color{black} In Fig. 3, it is demonstrated that as $\gamma$ increases, the efficiency of the QSHE gradually decreases and eventually levels off. This indicates that stronger spin-spin coupling significantly diminishes the QSHE's efficiency. Fig. 3(a) further reveals that an increase in $\gamma$ leads to a rise in the temperature of the cold reservoir ($T_C$), which causes the efficiency to drop faster. We also find in Figs.  3(a)–3(d) that the QSHE efficiency improves with an increasing parameter $\alpha$ when $\gamma$ is not too large. However, this effect is ultimately interrupted by the increase in $\gamma$. Regardless of the value of $\alpha$, the efficiency finally stabilizes at certain values as $\gamma$ continues to grow.}

   \begin{figure*}[htp]
	\centering
	\includegraphics[width=0.68\columnwidth]{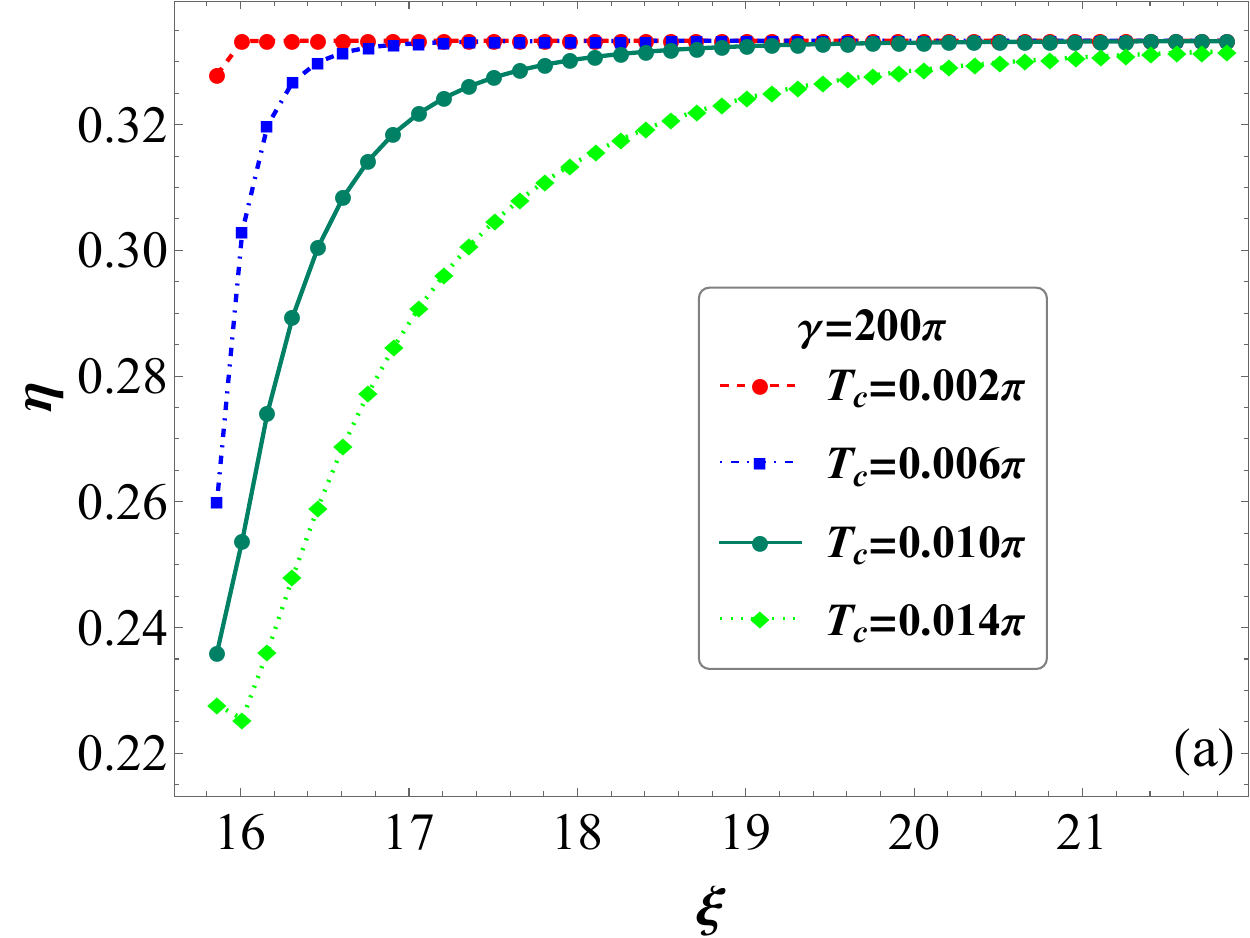}
	\includegraphics[width=0.68\columnwidth]{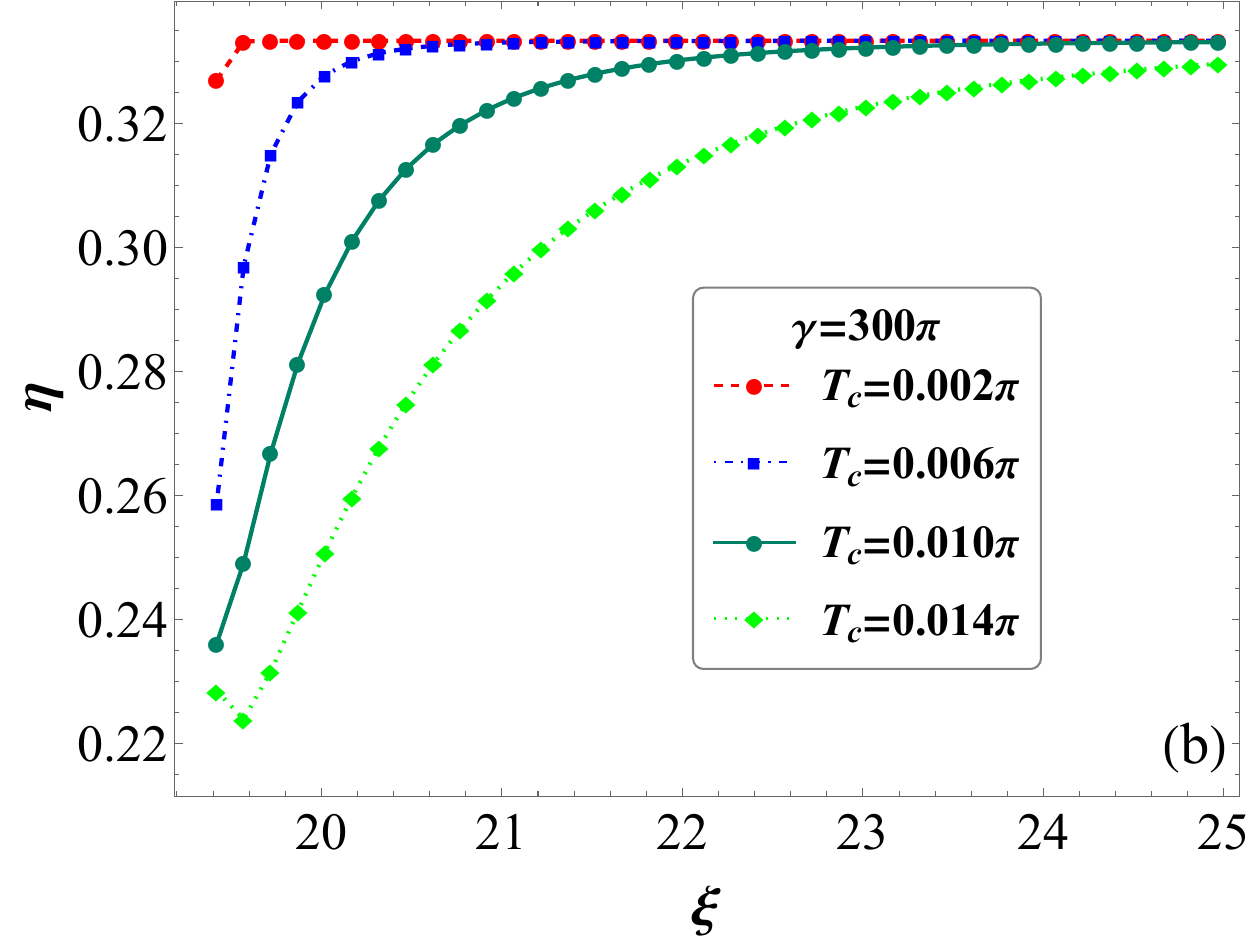}
	\includegraphics[width=0.68\columnwidth]{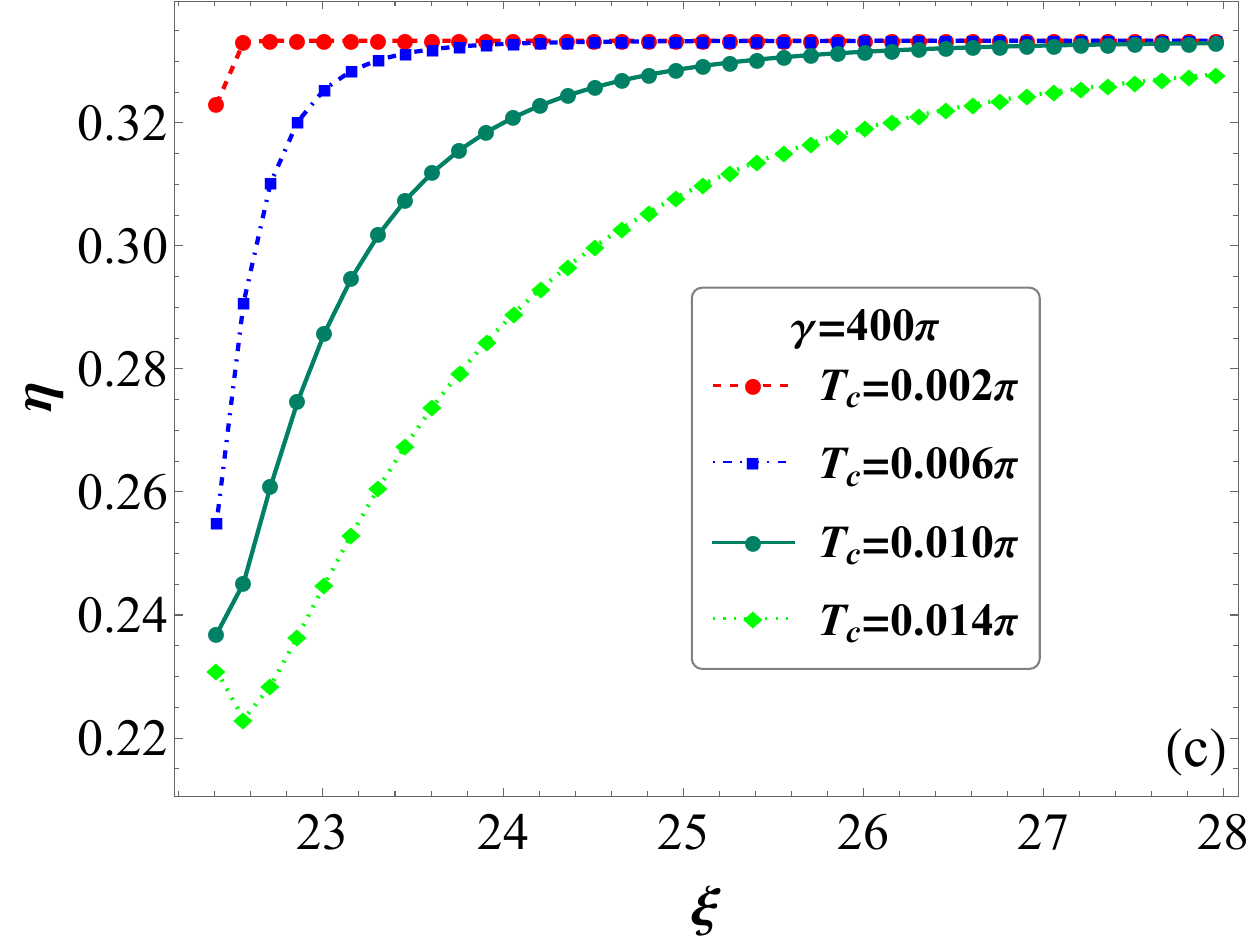}
    \includegraphics[width=0.68\columnwidth]{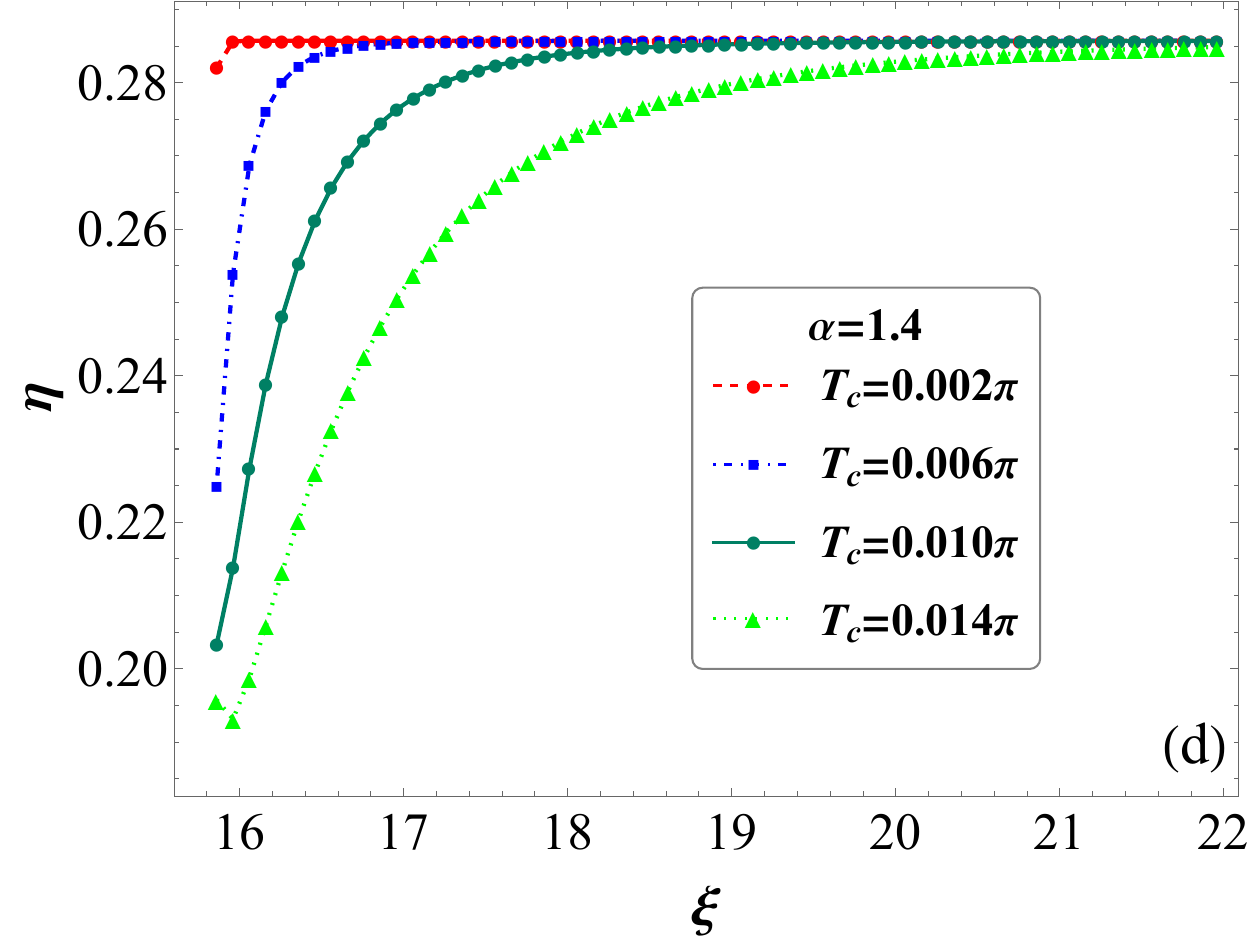}
	\includegraphics[width=0.68\columnwidth]{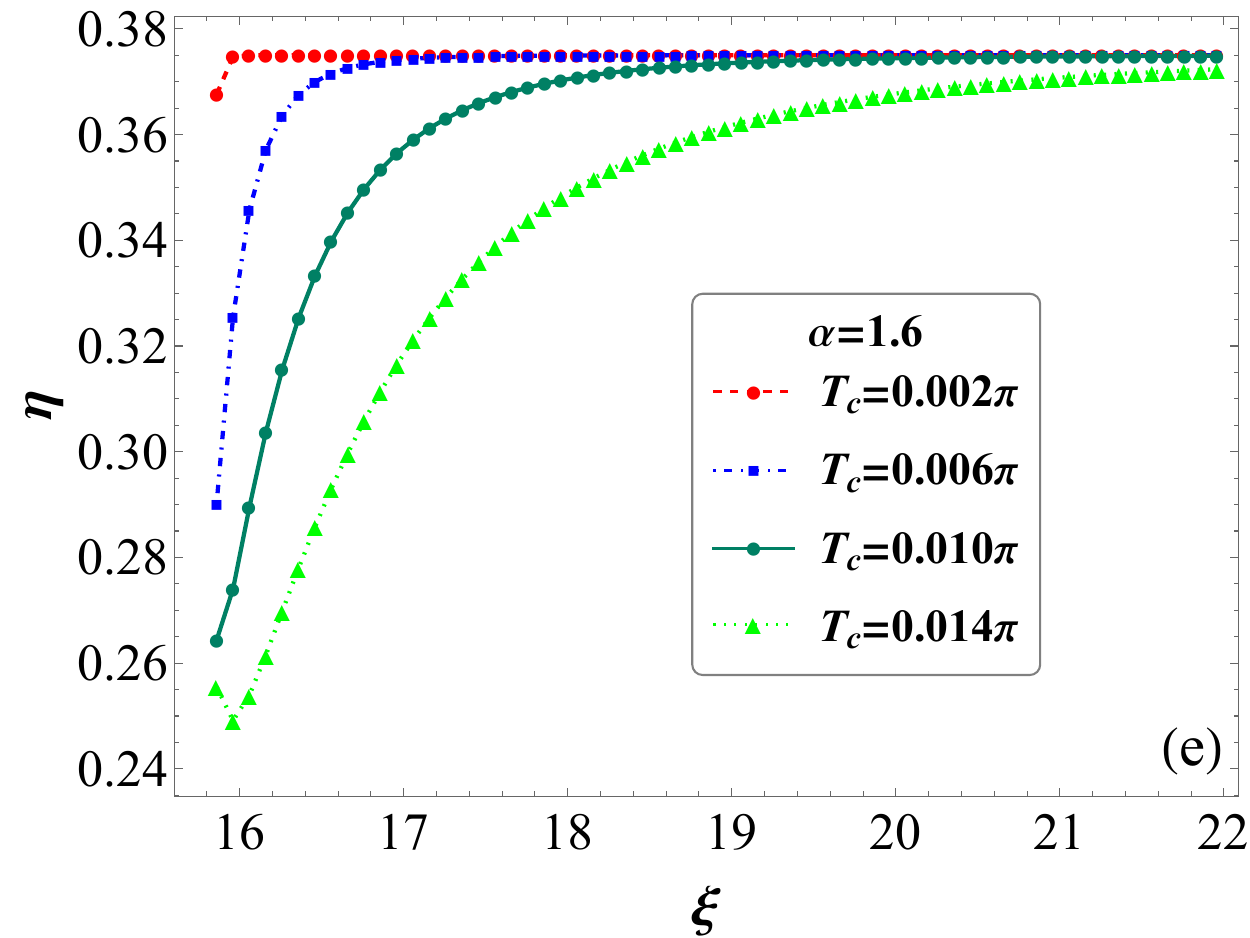}
	\includegraphics[width=0.68\columnwidth]{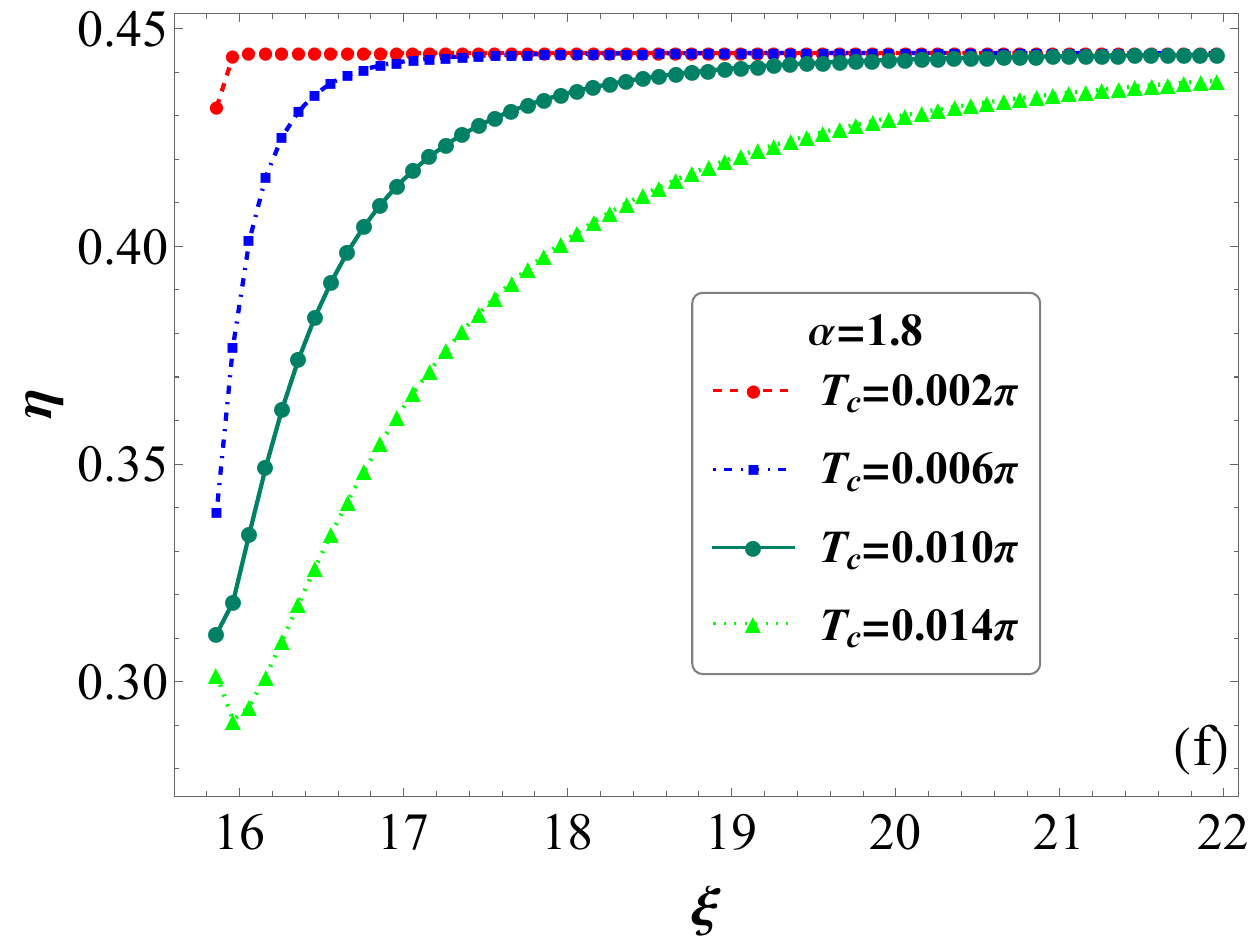}
	\caption{In the superradiative phase, the efficiency of QSHE $\eta$ as a function of $\xi$ at different temperatures $T_{c}=0.002 \pi$ (the red dotted line with solid circles), $T_{c}=0.006 \pi$ (the blue dotted line with solid squares), $T_{c}=0.010 \pi$ (the dark green solid line with solid circles), $T_{c}=0.014 \pi$ (the green dotted line with solid circles).We choose the coupling strength $\gamma=200 \pi$ for (a), $\gamma=300 \pi$ for (b), and $\gamma=400 \pi$ for (c), and $\gamma=200 \pi$ for (e)-(f). As the value of $\alpha$, it is $\alpha$= 1.5 for (a)-(c), $\alpha$ = 1.4 for (d), $\alpha$ = 1.6 for (e), and $\alpha$ = 1.8 for (f).}
	\label{wildlife}
    \end{figure*}

{\color{black} In the superradiant phase, we also explore the dependence of the QSHE efficiency on the parameter $\xi$ by varying system parameters. We summarize the numerical results in Fig. 4, illustrating the variation of the efficiency as a function of $\xi$. Specifically, we choose the parameter $\alpha$ to be constant and change $T_C$ near the critical point and the parameter $\gamma$.} {\color{black}It is demonstrated in Fig. 4(a)} that when $\gamma$ and $\alpha$ are constant, low $T_C$ results in {\color{black}slight changes in the efficiency with increasing $\xi$, and its value approaches the Carnot limit.} As $T_C$ increases, the efficiency {\color{black}first grows progressively with $\xi$, and eventually approaches a constant that is the Carnot limit.}
A detailed comparison of the efficiency curves at different temperatures reveals that the $\xi$ value required to reach {\color{black}the Carnot limit} increases with $T_C$. This {\color{black} finding highlights the covariant effect of the two parameters} $T_C$ and $\xi$ on the QSHE efficiency. Figs. 4(a)–4(c) show that for a fixed $\alpha$, a higher $\gamma$ also necessitates a larger $\xi$ to achieve the {\color{black}Carnot limit}. Similarly, Figs. 4(d)–4(e) {\color{black}indicate} that as $\gamma$ increases given a fixed $\alpha$, the $\xi$ value needed to reach this efficiency limit {\color{black}gets larger.}

   \section{Possible realization in an Ion Trap}
  {\color{black}In this section, we discuss the experimental realization of the system described by the two-qubit QRM with spin-spin coupling. The system interaction is possibly implementable in a physical system of trapped ions.
  Consider the ion-trap experimental system realized in Refs. \cite{Schneider_2012, Lemmer_2018},} two ions are confined within a linear Paul trap, where electric or magnetic fields are adjustable to form stable Coulomb crystals. These ions are then cooled to their kinematic ground state using Doppler and sideband cooling techniques. After cooling, the ion-ion interactions are primarily controlled by their charges and Coulomb forces. At low energies, the ions' motion can be described by a harmonic oscillator, and the two energy levels of their electronic ground state form a qubit. For example, with a \(^{171}\mathrm{Yb}\) ion, the qubit is represented by $\left | \downarrow  \right \rangle=\left | S_{1/2},F=0,m_{F}=0\right \rangle $ and $\left | \uparrow  \right \rangle=\left | S_{1/2},F=1,m_{F}=0\right \rangle$, with a transition frequency \(\varepsilon_i\) between the two energy levels. 
The system's total Hamiltonian consists of an uncoupled part and a coupled part. The uncoupled Hamiltonian is given by \(\hat{H}_s = \varepsilon_i \hat{\sigma}_i^z + \sum_{m=1}^{3N} \hbar \omega_m a_m^\dagger a_m\). The coupled Hamiltonian includes ion-ion interactions, represented by \(\hat{H}_{\text{ion-ion}} = J_{12} \hat{\sigma}_1^x \hat{\sigma}_2^x\), and ion-laser interactions, given by \(\hat{H}_{\text{ion-laser}} = \sum_{i=1}^{N=2} \hbar \Omega_I^{(i)} (\exp[i(\vec{k}_I^{(i)} \cdot \hat{\vec{r}}^{(i)} - \omega_I t + \phi_I^{(i)})] + \text{h.c.}) \hat{\sigma}_i^z\).
Here, the interaction strength between the two ions, \(J_{12}\), depends on the Rabi coupling strengths \(\Omega_I^{(1)}\) and \(\Omega_I^{(2)}\), and the phonon modes. The position operator \(\hat{\vec{r}}^{(i)}\) of each ion can be split into its equilibrium position and displacement. The resulting ion-laser {\color{black} interaction can be simplified using phonon operators and the Lamb-Dicke factor}, which measures the extent of ion displacement relative to the wavelength of the laser light. Thus, the ion-laser interaction Hamiltonian can be expressed as, 
$\hat{H}_{ion-laser}= \sum_{i=1}^{N=2}\hbar \Omega _{I}^{\left ( i \right ) }(\exp (i[ \sum_{m=1}^{3N}\eta_{im}(a_{m} ^{\dagger}+a_{m} )-\omega _{I} t +\phi _{I}^{(i ) }]) +\mathrm{h.c.})\hat{\sigma}_{i}^{z}$.

{\color{black}In the interaction picture with rotating wave approximation applied, the total Hamiltonian of the system is given by,}
\begin{equation}
	\begin{split}
		\hat{H^{'}}= &\hbar \varepsilon_{i} \hat{\sigma}_{i}^{z}+\omega _{0}\hat{a} ^{\dagger}\hat{a}+J_{12}\hat{\sigma} _{1}^{x}\hat{\sigma}_{2}^{x} \\
		&+i\hbar \Omega _{I}^{\left ( i \right ) }\sum_{m=1}^{3N}\eta_{m}^{\left ( i \right ) }e^{i(\delta_{m}t +\phi _{I}^{\left ( i \right ) })}a_{m} ^{\dagger}\hat{\sigma}_{i}^{z}+\mathrm{h.c.}\quad (i=1,2),\\		
	\end{split}
 \end{equation}
where $\delta_{m}: = \omega _{I} -\omega _{m}$. Furthermore, the Hamiltonian of the two-qubit QRM with spin-spin coupling is thus obtained, 
\begin{equation}
\tilde{H} = \omega_0 \hat{a}^\dagger \hat{a} + \sum_{i=1}^{n=2} \varepsilon_i \hat{\sigma}_i^z + \gamma \hat{\sigma}_1^x \hat{\sigma}_2^x + \sum_{i=1}^{n=2} \lambda_i \hat{\sigma}_i^z (\hat{a}^\dagger + \hat{a}),
\end{equation}
where \(\lambda_i = i\hbar \Omega_I^{(i)} \sum_{m=1}^{3N} \eta_{im}\), and \(\gamma = J_{12} = \Omega_1^{(1)} \Omega_2^{(2)} \sum_{m=1}^{3N} \frac{\eta_{1m} \eta_{2m} \omega_m}{\mu^2 - \omega_m^2}\).

By adjusting parameters like \(\Omega_I^{(i)}\), \(\eta_{im}\), and \(\omega_m\), we can control the strengths of the spin-spin and spin-mode interactions to achieve the desired results {\color{black} discussed in our scheme.}
  
  \section{Conclusion}

{\color{black}To summarize, we have explored a quantum Stirling cycle based on two-qubit QRM together with spin-spin coupling as the working medium. It has been shown that the efficiency of the QSHE can be improved via parameter optimization, considering the different ways that the effective coupling constant can approach its critical value. Specifically, we investigated the efficiency of the QSHE throughout its thermodynamic cycle by changing spin-spin coupling strength, spin-mode coupling strength, mode frequency, and temperature.}

When analyzing the thermodynamic properties of the quantum Stirling cycle {\color{black} in the normal phase}, it is necessary to consider its two isochoric processes that are characterized by the effective coupling constants $g_1$ and $g_2$. {\color{black} The effective coupling constants are influenced by system parameters.} By carefully adjusting these parameters to ensure $g_1 < g_2$, we have observed that an increase in the spin-spin coupling strength significantly diminishes the engine's efficiency. {\color{black} 
 Conversely, the efficiency can be enhanced through increasing the spin-mode coupling strength.} Near the critical point of the system and at low temperatures, the efficiency of the heat engine approaches the Carnot limit.

In the superradiant phase, {\color{black} we have found that the efficiency approaches the Carnot limit at low temperatures and grows with an increasing ratio of spin-mode coupling strength to mode frequency, providing a fixed ratio between the hot and cold reservoir temperatures.} For a constant spin-spin coupling strength, {\color{black} the Carnot efficiency is achievable given the large enough spin-mode coupling strength and the temperature ratio}.

This study provides a comprehensive analysis of the thermodynamic properties of the quantum Stirling cycle, particularly {\color{black}about the effect of different approaches moving} to the critical point on the performance of the engine. {\color{black}These insights can help to better understand the principles of QSHEs and present operative methods for the design and optimization of efficient QHEs, boosting further developments in this research area.}

	
 \section{Acknowledgment}
	C.R. was supported by the National Natural Science Foundation of China (Grants No. 12075245, 12421005 and No. 12247105), Hunan provincial major sci-tech program (No. 2023ZJ1010), the Natural Science Foundation of Hunan Province (2021JJ10033), the Foundation Xiangjiang Laboratory (XJ2302001) and Xiaoxiang Scholars Program of Hunan Normal University. C.W. is supported by the National Research Foundation, Singapore and A*STAR under its Quantum Engineering Programme (NRF2021-QEP2-02-P03).
	\bibliographystyle{apsrev4-1}
	\bibliography{ref}

\begin{thebibliography}{73}%
\makeatletter
\providecommand \@ifxundefined [1]{%
 \@ifx{#1\undefined}
}%
\providecommand \@ifnum [1]{%
 \ifnum #1\expandafter \@firstoftwo
 \else \expandafter \@secondoftwo
 \fi
}%
\providecommand \@ifx [1]{%
 \ifx #1\expandafter \@firstoftwo
 \else \expandafter \@secondoftwo
 \fi
}%
\providecommand \natexlab [1]{#1}%
\providecommand \enquote  [1]{``#1''}%
\providecommand \bibnamefont  [1]{#1}%
\providecommand \bibfnamefont [1]{#1}%
\providecommand \citenamefont [1]{#1}%
\providecommand \href@noop [0]{\@secondoftwo}%
\providecommand \href [0]{\begingroup \@sanitize@url \@href}%
\providecommand \@href[1]{\@@startlink{#1}\@@href}%
\providecommand \@@href[1]{\endgroup#1\@@endlink}%
\providecommand \@sanitize@url [0]{\catcode `\\12\catcode `\$12\catcode
  `\&12\catcode `\#12\catcode `\^12\catcode `\_12\catcode `\%12\relax}%
\providecommand \@@startlink[1]{}%
\providecommand \@@endlink[0]{}%
\providecommand \url  [0]{\begingroup\@sanitize@url \@url }%
\providecommand \@url [1]{\endgroup\@href {#1}{\urlprefix }}%
\providecommand \urlprefix  [0]{URL }%
\providecommand \Eprint [0]{\href }%
\providecommand \doibase [0]{http://dx.doi.org/}%
\providecommand \selectlanguage [0]{\@gobble}%
\providecommand \bibinfo  [0]{\@secondoftwo}%
\providecommand \bibfield  [0]{\@secondoftwo}%
\providecommand \translation [1]{[#1]}%
\providecommand \BibitemOpen [0]{}%
\providecommand \bibitemStop [0]{}%
\providecommand \bibitemNoStop [0]{.\EOS\space}%
\providecommand \EOS [0]{\spacefactor3000\relax}%
\providecommand \BibitemShut  [1]{\csname bibitem#1\endcsname}%
\let\auto@bib@innerbib\@empty
\bibitem [{\citenamefont {Scovil}\ and\ \citenamefont
  {Schulz-DuBois}(1959)}]{scovil1959three}%
  \BibitemOpen
  \bibfield  {author} {\bibinfo {author} {\bibfnamefont {H.~E.~D.}\
  \bibnamefont {Scovil}}\ and\ \bibinfo {author} {\bibfnamefont {E.~O.}\
  \bibnamefont {Schulz-DuBois}},\ }\href@noop {} {\bibfield  {journal}
  {\bibinfo  {journal} {Phys. Rev. Lett.}\ }\textbf {\bibinfo {volume} {2}},\
  \bibinfo {pages} {262} (\bibinfo {year} {1959})}\BibitemShut {NoStop}%
\bibitem [{\citenamefont {Kosloff}(1984)}]{kosloff1984quantum}%
  \BibitemOpen
  \bibfield  {author} {\bibinfo {author} {\bibfnamefont {R.}~\bibnamefont
  {Kosloff}},\ }\href@noop {} {\bibfield  {journal} {\bibinfo  {journal} {J.
  Chem. Phys.}\ }\textbf {\bibinfo {volume} {80}},\ \bibinfo {pages} {1625}
  (\bibinfo {year} {1984})}\BibitemShut {NoStop}%
\bibitem [{\citenamefont {Kosloff}(2013)}]{Kosloff2013Quantum}%
  \BibitemOpen
  \bibfield  {author} {\bibinfo {author} {\bibfnamefont {R.}~\bibnamefont
  {Kosloff}},\ }\href@noop {} {\bibfield  {journal} {\bibinfo  {journal}
  {Entropy}\ }\textbf {\bibinfo {volume} {15}},\ \bibinfo {pages} {2100}
  (\bibinfo {year} {2013})}\BibitemShut {NoStop}%
\bibitem [{\citenamefont {Goold}\ \emph {et~al.}(2016)\citenamefont {Goold},
  \citenamefont {Huber}, \citenamefont {Riera}, \citenamefont {Del~Rio},\ and\
  \citenamefont {Skrzypczyk}}]{goold2016d}%
  \BibitemOpen
  \bibfield  {author} {\bibinfo {author} {\bibfnamefont {J.}~\bibnamefont
  {Goold}}, \bibinfo {author} {\bibfnamefont {M.}~\bibnamefont {Huber}},
  \bibinfo {author} {\bibfnamefont {A.}~\bibnamefont {Riera}}, \bibinfo
  {author} {\bibfnamefont {L.}~\bibnamefont {Del~Rio}}, \ and\ \bibinfo
  {author} {\bibfnamefont {P.}~\bibnamefont {Skrzypczyk}},\ }\href@noop {}
  {\bibfield  {journal} {\bibinfo  {journal} {J. Phys. A: Math. Theor.}\
  }\textbf {\bibinfo {volume} {49}},\ \bibinfo {pages} {143001} (\bibinfo
  {year} {2016})}\BibitemShut {NoStop}%
\bibitem [{\citenamefont {Vinjanampathy}\ and\ \citenamefont
  {Anders}(2016)}]{Sai2016Quantum}%
  \BibitemOpen
  \bibfield  {author} {\bibinfo {author} {\bibfnamefont {S.}~\bibnamefont
  {Vinjanampathy}}\ and\ \bibinfo {author} {\bibfnamefont {J.}~\bibnamefont
  {Anders}},\ }\href@noop {} {\bibfield  {journal} {\bibinfo  {journal}
  {Contemp. Phys.}\ }\textbf {\bibinfo {volume} {57}},\ \bibinfo {pages} {545}
  (\bibinfo {year} {2016})}\BibitemShut {NoStop}%
\bibitem [{\citenamefont {Deffner}\ and\ \citenamefont
  {Campbell}(2019)}]{deffner2019quantum}%
  \BibitemOpen
  \bibfield  {author} {\bibinfo {author} {\bibfnamefont {S.}~\bibnamefont
  {Deffner}}\ and\ \bibinfo {author} {\bibfnamefont {S.}~\bibnamefont
  {Campbell}},\ }\href@noop {} {\emph {\bibinfo {title} {Quantum
  Thermodynamics: An introduction to the thermodynamics of quantum
  information}}}\ (\bibinfo  {publisher} {Morgan \& Claypool Publishers},\
  \bibinfo {year} {2019})\BibitemShut {NoStop}%
\bibitem [{\citenamefont {Quan}\ \emph {et~al.}(2007)\citenamefont {Quan},
  \citenamefont {Liu}, \citenamefont {Sun},\ and\ \citenamefont
  {Nori}}]{quan2007quantum}%
  \BibitemOpen
  \bibfield  {author} {\bibinfo {author} {\bibfnamefont {H.-T.}\ \bibnamefont
  {Quan}}, \bibinfo {author} {\bibfnamefont {Y.-X.}\ \bibnamefont {Liu}},
  \bibinfo {author} {\bibfnamefont {C.-P.}\ \bibnamefont {Sun}}, \ and\
  \bibinfo {author} {\bibfnamefont {F.}~\bibnamefont {Nori}},\ }\href@noop {}
  {\bibfield  {journal} {\bibinfo  {journal} {Phys. Rev. E}\ }\textbf {\bibinfo
  {volume} {76}},\ \bibinfo {pages} {031105} (\bibinfo {year}
  {2007})}\BibitemShut {NoStop}%
\bibitem [{\citenamefont {Scully}(2010)}]{scully2010quantum}%
  \BibitemOpen
  \bibfield  {author} {\bibinfo {author} {\bibfnamefont {M.~O.}\ \bibnamefont
  {Scully}},\ }\href@noop {} {\bibfield  {journal} {\bibinfo  {journal} {Phys.
  Rev. Lett.}\ }\textbf {\bibinfo {volume} {104}},\ \bibinfo {pages} {207701}
  (\bibinfo {year} {2010})}\BibitemShut {NoStop}%
\bibitem [{\citenamefont {Huang}\ \emph {et~al.}(2012)\citenamefont {Huang},
  \citenamefont {Wang},\ and\ \citenamefont {Yi}}]{PhysRevE.86.051105}%
  \BibitemOpen
  \bibfield  {author} {\bibinfo {author} {\bibfnamefont {X.}~\bibnamefont
  {Huang}}, \bibinfo {author} {\bibfnamefont {T.}~\bibnamefont {Wang}}, \ and\
  \bibinfo {author} {\bibfnamefont {X.}~\bibnamefont {Yi}},\ }\href@noop {}
  {\bibfield  {journal} {\bibinfo  {journal} {Phys. Rev. E}\ }\textbf {\bibinfo
  {volume} {86}},\ \bibinfo {pages} {051105} (\bibinfo {year}
  {2012})}\BibitemShut {NoStop}%
\bibitem [{\citenamefont {Scully}\ \emph {et~al.}(2011)\citenamefont {Scully},
  \citenamefont {Chapin}, \citenamefont {Dorfman}, \citenamefont {Kim},\ and\
  \citenamefont {Svidzinsky}}]{Scully2011QuantumHE}%
  \BibitemOpen
  \bibfield  {author} {\bibinfo {author} {\bibfnamefont {M.~O.}\ \bibnamefont
  {Scully}}, \bibinfo {author} {\bibfnamefont {K.~R.}\ \bibnamefont {Chapin}},
  \bibinfo {author} {\bibfnamefont {K.~E.}\ \bibnamefont {Dorfman}}, \bibinfo
  {author} {\bibfnamefont {M.~B.}\ \bibnamefont {Kim}}, \ and\ \bibinfo
  {author} {\bibfnamefont {A.}~\bibnamefont {Svidzinsky}},\ }\href@noop {}
  {\bibfield  {journal} {\bibinfo  {journal} {Proc. Natl. Acad. Sci. U.S.A.}\
  }\textbf {\bibinfo {volume} {108}},\ \bibinfo {pages} {15097 } (\bibinfo
  {year} {2011})}\BibitemShut {NoStop}%
\bibitem [{\citenamefont {Klatzow}\ \emph {et~al.}(2019)\citenamefont
  {Klatzow}, \citenamefont {Becker}, \citenamefont {Ledingham}, \citenamefont
  {Weinzetl}, \citenamefont {Kaczmarek}, \citenamefont {Saunders},
  \citenamefont {Nunn}, \citenamefont {Walmsley}, \citenamefont {Uzdin},\ and\
  \citenamefont {Poem}}]{PhysRevLett.122.110601}%
  \BibitemOpen
  \bibfield  {author} {\bibinfo {author} {\bibfnamefont {J.}~\bibnamefont
  {Klatzow}}, \bibinfo {author} {\bibfnamefont {J.~N.}\ \bibnamefont {Becker}},
  \bibinfo {author} {\bibfnamefont {P.~M.}\ \bibnamefont {Ledingham}}, \bibinfo
  {author} {\bibfnamefont {C.}~\bibnamefont {Weinzetl}}, \bibinfo {author}
  {\bibfnamefont {K.~T.}\ \bibnamefont {Kaczmarek}}, \bibinfo {author}
  {\bibfnamefont {D.~J.}\ \bibnamefont {Saunders}}, \bibinfo {author}
  {\bibfnamefont {J.}~\bibnamefont {Nunn}}, \bibinfo {author} {\bibfnamefont
  {I.~A.}\ \bibnamefont {Walmsley}}, \bibinfo {author} {\bibfnamefont
  {R.}~\bibnamefont {Uzdin}}, \ and\ \bibinfo {author} {\bibfnamefont
  {E.}~\bibnamefont {Poem}},\ }\href@noop {} {\bibfield  {journal} {\bibinfo
  {journal} {Phys. Rev. Lett.}\ }\textbf {\bibinfo {volume} {122}},\ \bibinfo
  {pages} {110601} (\bibinfo {year} {2019})}\BibitemShut {NoStop}%
\bibitem [{\citenamefont {Kammerlander}\ and\ \citenamefont
  {Anders}(2016)}]{Kammerlander2015CoherenceAM}%
  \BibitemOpen
  \bibfield  {author} {\bibinfo {author} {\bibfnamefont {P.}~\bibnamefont
  {Kammerlander}}\ and\ \bibinfo {author} {\bibfnamefont {J.}~\bibnamefont
  {Anders}},\ }\href@noop {} {\bibfield  {journal} {\bibinfo  {journal} {Sci.
  Rep}\ }\textbf {\bibinfo {volume} {6}},\ \bibinfo {pages} {1} (\bibinfo
  {year} {2016})}\BibitemShut {NoStop}%
\bibitem [{\citenamefont {Brandner}\ \emph {et~al.}(2017)\citenamefont
  {Brandner}, \citenamefont {Bauer},\ and\ \citenamefont
  {Seifert}}]{PhysRevLett.119.170602}%
  \BibitemOpen
  \bibfield  {author} {\bibinfo {author} {\bibfnamefont {K.}~\bibnamefont
  {Brandner}}, \bibinfo {author} {\bibfnamefont {M.}~\bibnamefont {Bauer}}, \
  and\ \bibinfo {author} {\bibfnamefont {U.}~\bibnamefont {Seifert}},\
  }\href@noop {} {\bibfield  {journal} {\bibinfo  {journal} {Phys. Rev. Lett.}\
  }\textbf {\bibinfo {volume} {119}},\ \bibinfo {pages} {170602} (\bibinfo
  {year} {2017})}\BibitemShut {NoStop}%
\bibitem [{\citenamefont {Dorfman}\ \emph {et~al.}(2018)\citenamefont
  {Dorfman}, \citenamefont {Xu},\ and\ \citenamefont
  {Cao}}]{Dorfman2018EfficiencyAM}%
  \BibitemOpen
  \bibfield  {author} {\bibinfo {author} {\bibfnamefont {K.~E.}\ \bibnamefont
  {Dorfman}}, \bibinfo {author} {\bibfnamefont {D.}~\bibnamefont {Xu}}, \ and\
  \bibinfo {author} {\bibfnamefont {J.}~\bibnamefont {Cao}},\ }\href@noop {}
  {\bibfield  {journal} {\bibinfo  {journal} {Phys. Rev. E}\ }\textbf {\bibinfo
  {volume} {97}},\ \bibinfo {pages} {042120} (\bibinfo {year}
  {2018})}\BibitemShut {NoStop}%
\bibitem [{\citenamefont {Dillenschneider}\ and\ \citenamefont
  {Lutz}(2009)}]{Dillenschneider_2009}%
  \BibitemOpen
  \bibfield  {author} {\bibinfo {author} {\bibfnamefont {R.}~\bibnamefont
  {Dillenschneider}}\ and\ \bibinfo {author} {\bibfnamefont {E.}~\bibnamefont
  {Lutz}},\ }\href@noop {} {\bibfield  {journal} {\bibinfo  {journal}
  {Europhys. Lett.}\ }\textbf {\bibinfo {volume} {88}},\ \bibinfo {pages}
  {50003} (\bibinfo {year} {2009})}\BibitemShut {NoStop}%
\bibitem [{\citenamefont {Abah}\ and\ \citenamefont {Lutz}(2014)}]{Abah_2014}%
  \BibitemOpen
  \bibfield  {author} {\bibinfo {author} {\bibfnamefont {O.}~\bibnamefont
  {Abah}}\ and\ \bibinfo {author} {\bibfnamefont {E.}~\bibnamefont {Lutz}},\
  }\href@noop {} {\bibfield  {journal} {\bibinfo  {journal} {Europhys. Lett.}\
  }\textbf {\bibinfo {volume} {106}},\ \bibinfo {pages} {20001} (\bibinfo
  {year} {2014})}\BibitemShut {NoStop}%
\bibitem [{\citenamefont {Niedenzu}\ \emph {et~al.}(2018)\citenamefont
  {Niedenzu}, \citenamefont {Mukherjee}, \citenamefont {Ghosh}, \citenamefont
  {Kofman},\ and\ \citenamefont {Kurizki}}]{Us_natcom}%
  \BibitemOpen
  \bibfield  {author} {\bibinfo {author} {\bibfnamefont {W.}~\bibnamefont
  {Niedenzu}}, \bibinfo {author} {\bibfnamefont {V.}~\bibnamefont {Mukherjee}},
  \bibinfo {author} {\bibfnamefont {A.}~\bibnamefont {Ghosh}}, \bibinfo
  {author} {\bibfnamefont {A.~G.}\ \bibnamefont {Kofman}}, \ and\ \bibinfo
  {author} {\bibfnamefont {G.}~\bibnamefont {Kurizki}},\ }\href@noop {}
  {\bibfield  {journal} {\bibinfo  {journal} {Nat. Commun.}\ }\textbf {\bibinfo
  {volume} {9}},\ \bibinfo {pages} {165} (\bibinfo {year} {2018})}\BibitemShut
  {NoStop}%
\bibitem [{\citenamefont {Klaers}\ \emph {et~al.}(2017)\citenamefont {Klaers},
  \citenamefont {Faelt}, \citenamefont {Imamoglu},\ and\ \citenamefont
  {Togan}}]{PhysRevX.7.031044}%
  \BibitemOpen
  \bibfield  {author} {\bibinfo {author} {\bibfnamefont {J.}~\bibnamefont
  {Klaers}}, \bibinfo {author} {\bibfnamefont {S.}~\bibnamefont {Faelt}},
  \bibinfo {author} {\bibfnamefont {A.}~\bibnamefont {Imamoglu}}, \ and\
  \bibinfo {author} {\bibfnamefont {E.}~\bibnamefont {Togan}},\ }\href@noop {}
  {\bibfield  {journal} {\bibinfo  {journal} {Phys. Rev. X}\ }\textbf {\bibinfo
  {volume} {7}},\ \bibinfo {pages} {031044} (\bibinfo {year}
  {2017})}\BibitemShut {NoStop}%
\bibitem [{\citenamefont {Ro{\ss}nagel}\ \emph {et~al.}(2014)\citenamefont
  {Ro{\ss}nagel}, \citenamefont {Abah}, \citenamefont {Schmidt-Kaler},
  \citenamefont {Singer},\ and\ \citenamefont {Lutz}}]{PhysRevLett.112.030602}%
  \BibitemOpen
  \bibfield  {author} {\bibinfo {author} {\bibfnamefont {J.}~\bibnamefont
  {Ro{\ss}nagel}}, \bibinfo {author} {\bibfnamefont {O.}~\bibnamefont {Abah}},
  \bibinfo {author} {\bibfnamefont {F.}~\bibnamefont {Schmidt-Kaler}}, \bibinfo
  {author} {\bibfnamefont {K.}~\bibnamefont {Singer}}, \ and\ \bibinfo {author}
  {\bibfnamefont {E.}~\bibnamefont {Lutz}},\ }\href@noop {} {\bibfield
  {journal} {\bibinfo  {journal} {Phys. Rev. Lett.}\ }\textbf {\bibinfo
  {volume} {112}},\ \bibinfo {pages} {030602} (\bibinfo {year}
  {2014})}\BibitemShut {NoStop}%
\bibitem [{\citenamefont {Manzano}\ \emph {et~al.}(2016)\citenamefont
  {Manzano}, \citenamefont {Galve}, \citenamefont {Zambrini},\ and\
  \citenamefont {Parrondo}}]{PhysRevE.93.052120}%
  \BibitemOpen
  \bibfield  {author} {\bibinfo {author} {\bibfnamefont {G.}~\bibnamefont
  {Manzano}}, \bibinfo {author} {\bibfnamefont {F.}~\bibnamefont {Galve}},
  \bibinfo {author} {\bibfnamefont {R.}~\bibnamefont {Zambrini}}, \ and\
  \bibinfo {author} {\bibfnamefont {J.~M.~R.}\ \bibnamefont {Parrondo}},\
  }\href@noop {} {\bibfield  {journal} {\bibinfo  {journal} {Phys. Rev. E}\
  }\textbf {\bibinfo {volume} {93}},\ \bibinfo {pages} {052120} (\bibinfo
  {year} {2016})}\BibitemShut {NoStop}%
\bibitem [{\citenamefont {Campisi}\ and\ \citenamefont {Fazio}(2016)}]{CF16}%
  \BibitemOpen
  \bibfield  {author} {\bibinfo {author} {\bibfnamefont {M.}~\bibnamefont
  {Campisi}}\ and\ \bibinfo {author} {\bibfnamefont {R.}~\bibnamefont
  {Fazio}},\ }\href@noop {} {\bibfield  {journal} {\bibinfo  {journal} {Nat.
  Commun.}\ }\textbf {\bibinfo {volume} {7}},\ \bibinfo {pages} {11895}
  (\bibinfo {year} {2016})}\BibitemShut {NoStop}%
\bibitem [{\citenamefont {Fogarty}\ and\ \citenamefont
  {Busch}(2020)}]{Fogarty_2021}%
  \BibitemOpen
  \bibfield  {author} {\bibinfo {author} {\bibfnamefont {T.}~\bibnamefont
  {Fogarty}}\ and\ \bibinfo {author} {\bibfnamefont {T.}~\bibnamefont
  {Busch}},\ }\href@noop {} {\bibfield  {journal} {\bibinfo  {journal} {Quantum
  Sci. Technol.}\ }\textbf {\bibinfo {volume} {6}},\ \bibinfo {pages} {015003}
  (\bibinfo {year} {2020})}\BibitemShut {NoStop}%
\bibitem [{\citenamefont {Ma}\ \emph {et~al.}(2017)\citenamefont {Ma},
  \citenamefont {Su},\ and\ \citenamefont {Sun}}]{PhysRevE.96.022143}%
  \BibitemOpen
  \bibfield  {author} {\bibinfo {author} {\bibfnamefont {Y.-H.}\ \bibnamefont
  {Ma}}, \bibinfo {author} {\bibfnamefont {S.-H.}\ \bibnamefont {Su}}, \ and\
  \bibinfo {author} {\bibfnamefont {C.-P.}\ \bibnamefont {Sun}},\ }\href@noop
  {} {\bibfield  {journal} {\bibinfo  {journal} {Phys. Rev. E}\ }\textbf
  {\bibinfo {volume} {96}},\ \bibinfo {pages} {022143} (\bibinfo {year}
  {2017})}\BibitemShut {NoStop}%
\bibitem [{\citenamefont {Niedenzu}\ and\ \citenamefont
  {Kurizki}(2018)}]{Niedenzu_2018}%
  \BibitemOpen
  \bibfield  {author} {\bibinfo {author} {\bibfnamefont {W.}~\bibnamefont
  {Niedenzu}}\ and\ \bibinfo {author} {\bibfnamefont {G.}~\bibnamefont
  {Kurizki}},\ }\href@noop {} {\bibfield  {journal} {\bibinfo  {journal} {New
  J. Phys.}\ }\textbf {\bibinfo {volume} {20}},\ \bibinfo {pages} {113038}
  (\bibinfo {year} {2018})}\BibitemShut {NoStop}%
\bibitem [{\citenamefont {Hardal}\ and\ \citenamefont
  {M{\"u}stecapl{\i}o{\u{g}}lu}(2015)}]{hardal2015superradiant}%
  \BibitemOpen
  \bibfield  {author} {\bibinfo {author} {\bibfnamefont {A.~{\"U}.}\
  \bibnamefont {Hardal}}\ and\ \bibinfo {author} {\bibfnamefont {{\"O}.~E.}\
  \bibnamefont {M{\"u}stecapl{\i}o{\u{g}}lu}},\ }\href@noop {} {\bibfield
  {journal} {\bibinfo  {journal} {Sci. Rep.}\ }\textbf {\bibinfo {volume}
  {5}},\ \bibinfo {pages} {12953} (\bibinfo {year} {2015})}\BibitemShut
  {NoStop}%
\bibitem [{\citenamefont {Yunger~Halpern}\ \emph {et~al.}(2019)\citenamefont
  {Yunger~Halpern}, \citenamefont {White}, \citenamefont {Gopalakrishnan},\
  and\ \citenamefont {Refael}}]{PhysRevB.99.024203}%
  \BibitemOpen
  \bibfield  {author} {\bibinfo {author} {\bibfnamefont {N.}~\bibnamefont
  {Yunger~Halpern}}, \bibinfo {author} {\bibfnamefont {C.~D.}\ \bibnamefont
  {White}}, \bibinfo {author} {\bibfnamefont {S.}~\bibnamefont
  {Gopalakrishnan}}, \ and\ \bibinfo {author} {\bibfnamefont {G.}~\bibnamefont
  {Refael}},\ }\href@noop {} {\bibfield  {journal} {\bibinfo  {journal} {Phys.
  Rev. B}\ }\textbf {\bibinfo {volume} {99}},\ \bibinfo {pages} {024203}
  (\bibinfo {year} {2019})}\BibitemShut {NoStop}%
\bibitem [{\citenamefont {Ma}\ \emph {et~al.}(2018)\citenamefont {Ma},
  \citenamefont {Xu}, \citenamefont {Dong},\ and\ \citenamefont
  {Sun}}]{PhysRevE.98.042112}%
  \BibitemOpen
  \bibfield  {author} {\bibinfo {author} {\bibfnamefont {Y.-H.}\ \bibnamefont
  {Ma}}, \bibinfo {author} {\bibfnamefont {D.}~\bibnamefont {Xu}}, \bibinfo
  {author} {\bibfnamefont {H.}~\bibnamefont {Dong}}, \ and\ \bibinfo {author}
  {\bibfnamefont {C.-P.}\ \bibnamefont {Sun}},\ }\href@noop {} {\bibfield
  {journal} {\bibinfo  {journal} {Phys. Rev. E}\ }\textbf {\bibinfo {volume}
  {98}},\ \bibinfo {pages} {042112} (\bibinfo {year} {2018})}\BibitemShut
  {NoStop}%
\bibitem [{\citenamefont {Fadaie}\ \emph {et~al.}(2018)\citenamefont {Fadaie},
  \citenamefont {Yunt},\ and\ \citenamefont
  {M{\"u}stecapl{\i}o{\u{g}}lu}}]{PhysRevE.98.052124}%
  \BibitemOpen
  \bibfield  {author} {\bibinfo {author} {\bibfnamefont {M.}~\bibnamefont
  {Fadaie}}, \bibinfo {author} {\bibfnamefont {E.}~\bibnamefont {Yunt}}, \ and\
  \bibinfo {author} {\bibfnamefont {{\"O}.~E.}\ \bibnamefont
  {M{\"u}stecapl{\i}o{\u{g}}lu}},\ }\href@noop {} {\bibfield  {journal}
  {\bibinfo  {journal} {Phys. Rev. E}\ }\textbf {\bibinfo {volume} {98}},\
  \bibinfo {pages} {052124} (\bibinfo {year} {2018})}\BibitemShut {NoStop}%
\bibitem [{\citenamefont {Abah}\ \emph {et~al.}(2012)\citenamefont {Abah},
  \citenamefont {Rossnagel}, \citenamefont {Jacob}, \citenamefont {Deffner},
  \citenamefont {Schmidt-Kaler}, \citenamefont {Singer},\ and\ \citenamefont
  {Lutz}}]{PhysRevLett.109.203006}%
  \BibitemOpen
  \bibfield  {author} {\bibinfo {author} {\bibfnamefont {O.}~\bibnamefont
  {Abah}}, \bibinfo {author} {\bibfnamefont {J.}~\bibnamefont {Rossnagel}},
  \bibinfo {author} {\bibfnamefont {G.}~\bibnamefont {Jacob}}, \bibinfo
  {author} {\bibfnamefont {S.}~\bibnamefont {Deffner}}, \bibinfo {author}
  {\bibfnamefont {F.}~\bibnamefont {Schmidt-Kaler}}, \bibinfo {author}
  {\bibfnamefont {K.}~\bibnamefont {Singer}}, \ and\ \bibinfo {author}
  {\bibfnamefont {E.}~\bibnamefont {Lutz}},\ }\href@noop {} {\bibfield
  {journal} {\bibinfo  {journal} {Phys. Rev. Lett.}\ }\textbf {\bibinfo
  {volume} {109}},\ \bibinfo {pages} {203006} (\bibinfo {year}
  {2012})}\BibitemShut {NoStop}%
\bibitem [{\citenamefont {Fialko}\ and\ \citenamefont
  {Hallwood}(2012)}]{PhysRevLett.108.085303}%
  \BibitemOpen
  \bibfield  {author} {\bibinfo {author} {\bibfnamefont {O.}~\bibnamefont
  {Fialko}}\ and\ \bibinfo {author} {\bibfnamefont {D.~W.}\ \bibnamefont
  {Hallwood}},\ }\href@noop {} {\bibfield  {journal} {\bibinfo  {journal}
  {Phys. Rev. Lett.}\ }\textbf {\bibinfo {volume} {108}},\ \bibinfo {pages}
  {085303} (\bibinfo {year} {2012})}\BibitemShut {NoStop}%
\bibitem [{\citenamefont {Brantut}\ \emph {et~al.}(2013)\citenamefont
  {Brantut}, \citenamefont {Grenier}, \citenamefont {Meineke}, \citenamefont
  {Stadler}, \citenamefont {Krinner}, \citenamefont {Kollath}, \citenamefont
  {Esslinger},\ and\ \citenamefont {Georges}}]{doi:10.1126/science.1242308}%
  \BibitemOpen
  \bibfield  {author} {\bibinfo {author} {\bibfnamefont {J.-P.}\ \bibnamefont
  {Brantut}}, \bibinfo {author} {\bibfnamefont {C.}~\bibnamefont {Grenier}},
  \bibinfo {author} {\bibfnamefont {J.}~\bibnamefont {Meineke}}, \bibinfo
  {author} {\bibfnamefont {D.}~\bibnamefont {Stadler}}, \bibinfo {author}
  {\bibfnamefont {S.}~\bibnamefont {Krinner}}, \bibinfo {author} {\bibfnamefont
  {C.}~\bibnamefont {Kollath}}, \bibinfo {author} {\bibfnamefont
  {T.}~\bibnamefont {Esslinger}}, \ and\ \bibinfo {author} {\bibfnamefont
  {A.}~\bibnamefont {Georges}},\ }\href@noop {} {\bibfield  {journal} {\bibinfo
   {journal} {Sci. Rep.}\ }\textbf {\bibinfo {volume} {342}},\ \bibinfo {pages}
  {713} (\bibinfo {year} {2013})}\BibitemShut {NoStop}%
\bibitem [{\citenamefont {Zhang}\ \emph {et~al.}(2014)\citenamefont {Zhang},
  \citenamefont {Bariani},\ and\ \citenamefont
  {Meystre}}]{PhysRevLett.112.150602}%
  \BibitemOpen
  \bibfield  {author} {\bibinfo {author} {\bibfnamefont {K.}~\bibnamefont
  {Zhang}}, \bibinfo {author} {\bibfnamefont {F.}~\bibnamefont {Bariani}}, \
  and\ \bibinfo {author} {\bibfnamefont {P.}~\bibnamefont {Meystre}},\
  }\href@noop {} {\bibfield  {journal} {\bibinfo  {journal} {Phys. Rev. Lett.}\
  }\textbf {\bibinfo {volume} {112}},\ \bibinfo {pages} {150602} (\bibinfo
  {year} {2014})}\BibitemShut {NoStop}%
\bibitem [{\citenamefont {Peterson}\ \emph {et~al.}(2019)\citenamefont
  {Peterson}, \citenamefont {Batalhao}, \citenamefont {Herrera}, \citenamefont
  {Souza}, \citenamefont {Sarthour}, \citenamefont {Oliveira},\ and\
  \citenamefont {Serra}}]{PhysRevLett.123.240601}%
  \BibitemOpen
  \bibfield  {author} {\bibinfo {author} {\bibfnamefont {J.~P.}\ \bibnamefont
  {Peterson}}, \bibinfo {author} {\bibfnamefont {T.~B.}\ \bibnamefont
  {Batalhao}}, \bibinfo {author} {\bibfnamefont {M.}~\bibnamefont {Herrera}},
  \bibinfo {author} {\bibfnamefont {A.~M.}\ \bibnamefont {Souza}}, \bibinfo
  {author} {\bibfnamefont {R.~S.}\ \bibnamefont {Sarthour}}, \bibinfo {author}
  {\bibfnamefont {I.~S.}\ \bibnamefont {Oliveira}}, \ and\ \bibinfo {author}
  {\bibfnamefont {R.~M.}\ \bibnamefont {Serra}},\ }\href@noop {} {\bibfield
  {journal} {\bibinfo  {journal} {Phys. Rev. Lett.}\ }\textbf {\bibinfo
  {volume} {123}},\ \bibinfo {pages} {240601} (\bibinfo {year}
  {2019})}\BibitemShut {NoStop}%
\bibitem [{\citenamefont {Guthrie}\ \emph {et~al.}(2022)\citenamefont
  {Guthrie}, \citenamefont {Satrya}, \citenamefont {Chang}, \citenamefont
  {Menczel}, \citenamefont {Nori},\ and\ \citenamefont
  {Pekola}}]{PhysRevApplied.17.064022}%
  \BibitemOpen
  \bibfield  {author} {\bibinfo {author} {\bibfnamefont {A.}~\bibnamefont
  {Guthrie}}, \bibinfo {author} {\bibfnamefont {C.~D.}\ \bibnamefont {Satrya}},
  \bibinfo {author} {\bibfnamefont {Y.-C.}\ \bibnamefont {Chang}}, \bibinfo
  {author} {\bibfnamefont {P.}~\bibnamefont {Menczel}}, \bibinfo {author}
  {\bibfnamefont {F.}~\bibnamefont {Nori}}, \ and\ \bibinfo {author}
  {\bibfnamefont {J.~P.}\ \bibnamefont {Pekola}},\ }\href@noop {} {\bibfield
  {journal} {\bibinfo  {journal} {Phys. Rev. Appl.}\ }\textbf {\bibinfo
  {volume} {17}},\ \bibinfo {pages} {064022} (\bibinfo {year}
  {2022})}\BibitemShut {NoStop}%
\bibitem [{\citenamefont {Ronzani}\ \emph {et~al.}(2018)\citenamefont
  {Ronzani}, \citenamefont {Karimi}, \citenamefont {Senior}, \citenamefont
  {Chang}, \citenamefont {Peltonen}, \citenamefont {Chen},\ and\ \citenamefont
  {Pekola}}]{Ronzani2018TunablePH}%
  \BibitemOpen
  \bibfield  {author} {\bibinfo {author} {\bibfnamefont {A.}~\bibnamefont
  {Ronzani}}, \bibinfo {author} {\bibfnamefont {B.}~\bibnamefont {Karimi}},
  \bibinfo {author} {\bibfnamefont {J.}~\bibnamefont {Senior}}, \bibinfo
  {author} {\bibfnamefont {Y.-C.}\ \bibnamefont {Chang}}, \bibinfo {author}
  {\bibfnamefont {J.~T.}\ \bibnamefont {Peltonen}}, \bibinfo {author}
  {\bibfnamefont {C.}~\bibnamefont {Chen}}, \ and\ \bibinfo {author}
  {\bibfnamefont {J.~P.}\ \bibnamefont {Pekola}},\ }\href@noop {} {\bibfield
  {journal} {\bibinfo  {journal} {Nat. Phys.}\ }\textbf {\bibinfo {volume}
  {14}},\ \bibinfo {pages} {991} (\bibinfo {year} {2018})}\BibitemShut
  {NoStop}%
\bibitem [{\citenamefont {Pekola}\ and\ \citenamefont
  {Khaymovich}(2019)}]{pekola2019thermodynamics}%
  \BibitemOpen
  \bibfield  {author} {\bibinfo {author} {\bibfnamefont {J.~P.}\ \bibnamefont
  {Pekola}}\ and\ \bibinfo {author} {\bibfnamefont {I.~M.}\ \bibnamefont
  {Khaymovich}},\ }\href@noop {} {\bibfield  {journal} {\bibinfo  {journal}
  {Annu. Rev. Condens. Matter Phys.}\ }\textbf {\bibinfo {volume} {10}},\
  \bibinfo {pages} {193} (\bibinfo {year} {2019})}\BibitemShut {NoStop}%
\bibitem [{\citenamefont {Chand}\ and\ \citenamefont
  {Biswas}(2018)}]{PhysRevE.98.052147}%
  \BibitemOpen
  \bibfield  {author} {\bibinfo {author} {\bibfnamefont {S.}~\bibnamefont
  {Chand}}\ and\ \bibinfo {author} {\bibfnamefont {A.}~\bibnamefont {Biswas}},\
  }\href@noop {} {\bibfield  {journal} {\bibinfo  {journal} {Phys. Rev. E}\
  }\textbf {\bibinfo {volume} {98}},\ \bibinfo {pages} {052147} (\bibinfo
  {year} {2018})}\BibitemShut {NoStop}%
\bibitem [{\citenamefont {B.~S}\ \emph {et~al.}(2020)\citenamefont {B.~S},
  \citenamefont {Mukherjee}, \citenamefont {Divakaran},\ and\ \citenamefont
  {del Campo}}]{PhysRevResearch.2.043247}%
  \BibitemOpen
  \bibfield  {author} {\bibinfo {author} {\bibfnamefont {R.}~\bibnamefont
  {B.~S}}, \bibinfo {author} {\bibfnamefont {V.}~\bibnamefont {Mukherjee}},
  \bibinfo {author} {\bibfnamefont {U.}~\bibnamefont {Divakaran}}, \ and\
  \bibinfo {author} {\bibfnamefont {A.}~\bibnamefont {del Campo}},\ }\href@noop
  {} {\bibfield  {journal} {\bibinfo  {journal} {Phys. Rev. Res.}\ }\textbf
  {\bibinfo {volume} {2}},\ \bibinfo {pages} {043247} (\bibinfo {year}
  {2020})}\BibitemShut {NoStop}%
\bibitem [{\citenamefont {Purkait}\ and\ \citenamefont
  {Biswas}(2022)}]{purkait2022performance}%
  \BibitemOpen
  \bibfield  {author} {\bibinfo {author} {\bibfnamefont {C.}~\bibnamefont
  {Purkait}}\ and\ \bibinfo {author} {\bibfnamefont {A.}~\bibnamefont
  {Biswas}},\ }\href@noop {} {\bibfield  {journal} {\bibinfo  {journal} {Phys.
  Lett. A}\ }\textbf {\bibinfo {volume} {442}},\ \bibinfo {pages} {128180}
  (\bibinfo {year} {2022})}\BibitemShut {NoStop}%
\bibitem [{\citenamefont {Alcalde}\ and\ \citenamefont
  {Arias}(2019)}]{alcalde2019quantum}%
  \BibitemOpen
  \bibfield  {author} {\bibinfo {author} {\bibfnamefont {M.~A.}\ \bibnamefont
  {Alcalde}}\ and\ \bibinfo {author} {\bibfnamefont {E.}~\bibnamefont
  {Arias}},\ }\href@noop {} {\bibfield  {journal} {\bibinfo  {journal}
  {arXiv:1906.00292}\ } (\bibinfo {year} {2019})}\BibitemShut {NoStop}%
\bibitem [{\citenamefont {Wang}\ \emph {et~al.}(2024)\citenamefont {Wang},
  \citenamefont {Yung}, \citenamefont {Xu}, \citenamefont {Liu},\ and\
  \citenamefont {Chen}}]{PhysRevA.109.022208}%
  \BibitemOpen
  \bibfield  {author} {\bibinfo {author} {\bibfnamefont {Y.-S.}\ \bibnamefont
  {Wang}}, \bibinfo {author} {\bibfnamefont {M.-H.}\ \bibnamefont {Yung}},
  \bibinfo {author} {\bibfnamefont {D.}~\bibnamefont {Xu}}, \bibinfo {author}
  {\bibfnamefont {M.}~\bibnamefont {Liu}}, \ and\ \bibinfo {author}
  {\bibfnamefont {X.}~\bibnamefont {Chen}},\ }\href@noop {} {\bibfield
  {journal} {\bibinfo  {journal} {Phys. Rev. A}\ }\textbf {\bibinfo {volume}
  {109}},\ \bibinfo {pages} {022208} (\bibinfo {year} {2024})}\BibitemShut
  {NoStop}%
\bibitem [{\citenamefont {Chen}\ \emph {et~al.}(2019)\citenamefont {Chen},
  \citenamefont {Watanabe}, \citenamefont {Yu}, \citenamefont {Guan},\ and\
  \citenamefont {del Campo}}]{chen2019interaction}%
  \BibitemOpen
  \bibfield  {author} {\bibinfo {author} {\bibfnamefont {Y.-Y.}\ \bibnamefont
  {Chen}}, \bibinfo {author} {\bibfnamefont {G.}~\bibnamefont {Watanabe}},
  \bibinfo {author} {\bibfnamefont {Y.-C.}\ \bibnamefont {Yu}}, \bibinfo
  {author} {\bibfnamefont {X.-W.}\ \bibnamefont {Guan}}, \ and\ \bibinfo
  {author} {\bibfnamefont {A.}~\bibnamefont {del Campo}},\ }\href@noop {}
  {\bibfield  {journal} {\bibinfo  {journal} {Npj Quantum Inf.}\ }\textbf
  {\bibinfo {volume} {5}},\ \bibinfo {pages} {88} (\bibinfo {year}
  {2019})}\BibitemShut {NoStop}%
\bibitem [{\citenamefont {Fusco}\ \emph {et~al.}(2016)\citenamefont {Fusco},
  \citenamefont {Paternostro},\ and\ \citenamefont
  {De~Chiara}}]{PhysRevE.94.052122}%
  \BibitemOpen
  \bibfield  {author} {\bibinfo {author} {\bibfnamefont {L.}~\bibnamefont
  {Fusco}}, \bibinfo {author} {\bibfnamefont {M.}~\bibnamefont {Paternostro}},
  \ and\ \bibinfo {author} {\bibfnamefont {G.}~\bibnamefont {De~Chiara}},\
  }\href@noop {} {\bibfield  {journal} {\bibinfo  {journal} {Phys. Rev. E}\
  }\textbf {\bibinfo {volume} {94}},\ \bibinfo {pages} {052122} (\bibinfo
  {year} {2016})}\BibitemShut {NoStop}%
\bibitem [{\citenamefont {Kloc}\ \emph {et~al.}(2019)\citenamefont {Kloc},
  \citenamefont {Cejnar},\ and\ \citenamefont {Schaller}}]{kloc2019collective}%
  \BibitemOpen
  \bibfield  {author} {\bibinfo {author} {\bibfnamefont {M.}~\bibnamefont
  {Kloc}}, \bibinfo {author} {\bibfnamefont {P.}~\bibnamefont {Cejnar}}, \ and\
  \bibinfo {author} {\bibfnamefont {G.}~\bibnamefont {Schaller}},\ }\href@noop
  {} {\bibfield  {journal} {\bibinfo  {journal} {Phys. Rev. E}\ }\textbf
  {\bibinfo {volume} {100}},\ \bibinfo {pages} {042126} (\bibinfo {year}
  {2019})}\BibitemShut {NoStop}%
\bibitem [{\citenamefont {Piccitto}\ \emph {et~al.}(2022)\citenamefont
  {Piccitto}, \citenamefont {Campisi},\ and\ \citenamefont
  {Rossini}}]{Piccitto_2022}%
  \BibitemOpen
  \bibfield  {author} {\bibinfo {author} {\bibfnamefont {G.}~\bibnamefont
  {Piccitto}}, \bibinfo {author} {\bibfnamefont {M.}~\bibnamefont {Campisi}}, \
  and\ \bibinfo {author} {\bibfnamefont {D.}~\bibnamefont {Rossini}},\
  }\href@noop {} {\bibfield  {journal} {\bibinfo  {journal} {New J. Phys.}\
  }\textbf {\bibinfo {volume} {24}},\ \bibinfo {pages} {103023} (\bibinfo
  {year} {2022})}\BibitemShut {NoStop}%
\bibitem [{\citenamefont {Mukherjee}\ \emph {et~al.}(2024)\citenamefont
  {Mukherjee}, \citenamefont {Divakaran} \emph {et~al.}}]{revathy2024quantum}%
  \BibitemOpen
  \bibfield  {author} {\bibinfo {author} {\bibfnamefont {V.}~\bibnamefont
  {Mukherjee}}, \bibinfo {author} {\bibfnamefont {U.}~\bibnamefont
  {Divakaran}},  \emph {et~al.},\ }\href@noop {} {\bibfield  {journal}
  {\bibinfo  {journal} {Eur. Phys. J. B}\ }\textbf {\bibinfo {volume} {97}},\
  \bibinfo {pages} {76} (\bibinfo {year} {2024})}\BibitemShut {NoStop}%
\bibitem [{\citenamefont {BS}\ \emph {et~al.}(2022)\citenamefont {BS},
  \citenamefont {Mukherjee},\ and\ \citenamefont {Divakaran}}]{bs2022bath}%
  \BibitemOpen
  \bibfield  {author} {\bibinfo {author} {\bibfnamefont {R.}~\bibnamefont
  {BS}}, \bibinfo {author} {\bibfnamefont {V.}~\bibnamefont {Mukherjee}}, \
  and\ \bibinfo {author} {\bibfnamefont {U.}~\bibnamefont {Divakaran}},\
  }\href@noop {} {\bibfield  {journal} {\bibinfo  {journal} {Entropy}\ }\textbf
  {\bibinfo {volume} {24}},\ \bibinfo {pages} {1458} (\bibinfo {year}
  {2022})}\BibitemShut {NoStop}%
\bibitem [{\citenamefont {Jussiau}\ \emph {et~al.}(2023)\citenamefont
  {Jussiau}, \citenamefont {Bresque}, \citenamefont {Auff{\`e}ves},
  \citenamefont {Murch},\ and\ \citenamefont {Jordan}}]{jussiau2023many}%
  \BibitemOpen
  \bibfield  {author} {\bibinfo {author} {\bibfnamefont {{\'E}.}~\bibnamefont
  {Jussiau}}, \bibinfo {author} {\bibfnamefont {L.}~\bibnamefont {Bresque}},
  \bibinfo {author} {\bibfnamefont {A.}~\bibnamefont {Auff{\`e}ves}}, \bibinfo
  {author} {\bibfnamefont {K.~W.}\ \bibnamefont {Murch}}, \ and\ \bibinfo
  {author} {\bibfnamefont {A.~N.}\ \bibnamefont {Jordan}},\ }\href@noop {}
  {\bibfield  {journal} {\bibinfo  {journal} {Phys. Rev. Res.}\ }\textbf
  {\bibinfo {volume} {5}},\ \bibinfo {pages} {033122} (\bibinfo {year}
  {2023})}\BibitemShut {NoStop}%
\bibitem [{\citenamefont {Dann}\ and\ \citenamefont
  {Kosloff}(2020{\natexlab{a}})}]{Dann_2020}%
  \BibitemOpen
  \bibfield  {author} {\bibinfo {author} {\bibfnamefont {R.}~\bibnamefont
  {Dann}}\ and\ \bibinfo {author} {\bibfnamefont {R.}~\bibnamefont {Kosloff}},\
  }\href@noop {} {\bibfield  {journal} {\bibinfo  {journal} {New J. Phys.}\
  }\textbf {\bibinfo {volume} {22}},\ \bibinfo {pages} {013055} (\bibinfo
  {year} {2020}{\natexlab{a}})}\BibitemShut {NoStop}%
\bibitem [{\citenamefont {Bender}\ \emph {et~al.}(2000)\citenamefont {Bender},
  \citenamefont {Brody},\ and\ \citenamefont {Meister}}]{bender2000quantum}%
  \BibitemOpen
  \bibfield  {author} {\bibinfo {author} {\bibfnamefont {C.~M.}\ \bibnamefont
  {Bender}}, \bibinfo {author} {\bibfnamefont {D.~C.}\ \bibnamefont {Brody}}, \
  and\ \bibinfo {author} {\bibfnamefont {B.~K.}\ \bibnamefont {Meister}},\
  }\href@noop {} {\bibfield  {journal} {\bibinfo  {journal} {J. Phys. A: Math.
  Gen.}\ }\textbf {\bibinfo {volume} {33}},\ \bibinfo {pages} {4427} (\bibinfo
  {year} {2000})}\BibitemShut {NoStop}%
\bibitem [{\citenamefont {Denzler}\ and\ \citenamefont
  {Lutz}(2021)}]{denzler2021power}%
  \BibitemOpen
  \bibfield  {author} {\bibinfo {author} {\bibfnamefont {T.}~\bibnamefont
  {Denzler}}\ and\ \bibinfo {author} {\bibfnamefont {E.}~\bibnamefont {Lutz}},\
  }\href@noop {} {\bibfield  {journal} {\bibinfo  {journal} {Phys. Rev. Res.}\
  }\textbf {\bibinfo {volume} {3}},\ \bibinfo {pages} {L032041} (\bibinfo
  {year} {2021})}\BibitemShut {NoStop}%
\bibitem [{\citenamefont {Sutantyo}(2020)}]{sutantyo2020three}%
  \BibitemOpen
  \bibfield  {author} {\bibinfo {author} {\bibfnamefont {T.~E.~P.}\
  \bibnamefont {Sutantyo}},\ }\href@noop {} {\bibfield  {journal} {\bibinfo
  {journal} {Jurnal Fisika Unand}\ }\textbf {\bibinfo {volume} {9}},\ \bibinfo
  {pages} {142} (\bibinfo {year} {2020})}\BibitemShut {NoStop}%
\bibitem [{\citenamefont {{\c{C}}akmak}\ \emph {et~al.}(2020)\citenamefont
  {{\c{C}}akmak}, \citenamefont {{\c{C}}and{\i}r},\ and\ \citenamefont
  {Altintas}}]{ccakmak2020construction}%
  \BibitemOpen
  \bibfield  {author} {\bibinfo {author} {\bibfnamefont {S.}~\bibnamefont
  {{\c{C}}akmak}}, \bibinfo {author} {\bibfnamefont {M.}~\bibnamefont
  {{\c{C}}and{\i}r}}, \ and\ \bibinfo {author} {\bibfnamefont {F.}~\bibnamefont
  {Altintas}},\ }\href@noop {} {\bibfield  {journal} {\bibinfo  {journal}
  {Quantum Inf. Process.}\ }\textbf {\bibinfo {volume} {19}},\ \bibinfo {pages}
  {314} (\bibinfo {year} {2020})}\BibitemShut {NoStop}%
\bibitem [{\citenamefont {Dann}\ and\ \citenamefont
  {Kosloff}(2020{\natexlab{b}})}]{dann2020quantum}%
  \BibitemOpen
  \bibfield  {author} {\bibinfo {author} {\bibfnamefont {R.}~\bibnamefont
  {Dann}}\ and\ \bibinfo {author} {\bibfnamefont {R.}~\bibnamefont {Kosloff}},\
  }\href@noop {} {\bibfield  {journal} {\bibinfo  {journal} {New J. Phys.}\
  }\textbf {\bibinfo {volume} {22}},\ \bibinfo {pages} {013055} (\bibinfo
  {year} {2020}{\natexlab{b}})}\BibitemShut {NoStop}%
\bibitem [{\citenamefont {Qiu}\ \emph {et~al.}(2020)\citenamefont {Qiu},
  \citenamefont {Fei}, \citenamefont {Pan},\ and\ \citenamefont
  {Quan}}]{qiu2020quantum}%
  \BibitemOpen
  \bibfield  {author} {\bibinfo {author} {\bibfnamefont {T.}~\bibnamefont
  {Qiu}}, \bibinfo {author} {\bibfnamefont {Z.}~\bibnamefont {Fei}}, \bibinfo
  {author} {\bibfnamefont {R.}~\bibnamefont {Pan}}, \ and\ \bibinfo {author}
  {\bibfnamefont {H.}~\bibnamefont {Quan}},\ }\href@noop {} {\bibfield
  {journal} {\bibinfo  {journal} {Phys. Rev. E}\ }\textbf {\bibinfo {volume}
  {101}},\ \bibinfo {pages} {032113} (\bibinfo {year} {2020})}\BibitemShut
  {NoStop}%
\bibitem [{\citenamefont {Sutantyo}\ \emph {et~al.}(2015)\citenamefont
  {Sutantyo}, \citenamefont {Belfaqih},\ and\ \citenamefont
  {Prayitno}}]{qstac635abib41}%
  \BibitemOpen
  \bibfield  {author} {\bibinfo {author} {\bibfnamefont {T.~E.~P.}\
  \bibnamefont {Sutantyo}}, \bibinfo {author} {\bibfnamefont {I.~H.}\
  \bibnamefont {Belfaqih}}, \ and\ \bibinfo {author} {\bibfnamefont
  {T.}~\bibnamefont {Prayitno}},\ }\href@noop {} {\bibfield  {journal}
  {\bibinfo  {journal} {AIP Conf. Proc.}\ }\textbf {\bibinfo {volume} {1677}},\
  \bibinfo {pages} {040011} (\bibinfo {year} {2015})}\BibitemShut {NoStop}%
\bibitem [{\citenamefont {Bengtsson}\ \emph {et~al.}(2018)\citenamefont
  {Bengtsson}, \citenamefont {Tengstrand}, \citenamefont {Wacker},
  \citenamefont {Samuelsson}, \citenamefont {Ueda}, \citenamefont {Linke},\
  and\ \citenamefont {Reimann}}]{PhysRevLett.120.100601}%
  \BibitemOpen
  \bibfield  {author} {\bibinfo {author} {\bibfnamefont {J.}~\bibnamefont
  {Bengtsson}}, \bibinfo {author} {\bibfnamefont {M.~N.}\ \bibnamefont
  {Tengstrand}}, \bibinfo {author} {\bibfnamefont {A.}~\bibnamefont {Wacker}},
  \bibinfo {author} {\bibfnamefont {P.}~\bibnamefont {Samuelsson}}, \bibinfo
  {author} {\bibfnamefont {M.}~\bibnamefont {Ueda}}, \bibinfo {author}
  {\bibfnamefont {H.}~\bibnamefont {Linke}}, \ and\ \bibinfo {author}
  {\bibfnamefont {S.}~\bibnamefont {Reimann}},\ }\href@noop {} {\bibfield
  {journal} {\bibinfo  {journal} {Phys. Rev. Lett.}\ }\textbf {\bibinfo
  {volume} {120}},\ \bibinfo {pages} {100601} (\bibinfo {year}
  {2018})}\BibitemShut {NoStop}%
\bibitem [{\citenamefont {Kim}\ \emph {et~al.}(2011)\citenamefont {Kim},
  \citenamefont {Sagawa}, \citenamefont {De~Liberato},\ and\ \citenamefont
  {Ueda}}]{PhysRevLett.106.070401}%
  \BibitemOpen
  \bibfield  {author} {\bibinfo {author} {\bibfnamefont {S.~W.}\ \bibnamefont
  {Kim}}, \bibinfo {author} {\bibfnamefont {T.}~\bibnamefont {Sagawa}},
  \bibinfo {author} {\bibfnamefont {S.}~\bibnamefont {De~Liberato}}, \ and\
  \bibinfo {author} {\bibfnamefont {M.}~\bibnamefont {Ueda}},\ }\href@noop {}
  {\bibfield  {journal} {\bibinfo  {journal} {Phys. Rev. Lett.}\ }\textbf
  {\bibinfo {volume} {106}},\ \bibinfo {pages} {070401} (\bibinfo {year}
  {2011})}\BibitemShut {NoStop}%
\bibitem [{\citenamefont {Hwang}\ \emph {et~al.}(2015)\citenamefont {Hwang},
  \citenamefont {Puebla},\ and\ \citenamefont
  {Plenio}}]{PhysRevLett.115.180404}%
  \BibitemOpen
  \bibfield  {author} {\bibinfo {author} {\bibfnamefont {M.-J.}\ \bibnamefont
  {Hwang}}, \bibinfo {author} {\bibfnamefont {R.}~\bibnamefont {Puebla}}, \
  and\ \bibinfo {author} {\bibfnamefont {M.~B.}\ \bibnamefont {Plenio}},\
  }\href@noop {} {\bibfield  {journal} {\bibinfo  {journal} {Phys. Rev. Lett.}\
  }\textbf {\bibinfo {volume} {115}},\ \bibinfo {pages} {180404} (\bibinfo
  {year} {2015})}\BibitemShut {NoStop}%
\bibitem [{\citenamefont {Puebla}\ \emph {et~al.}(2017)\citenamefont {Puebla},
  \citenamefont {Hwang}, \citenamefont {Casanova},\ and\ \citenamefont
  {Plenio}}]{PhysRevLett.118.073001}%
  \BibitemOpen
  \bibfield  {author} {\bibinfo {author} {\bibfnamefont {R.}~\bibnamefont
  {Puebla}}, \bibinfo {author} {\bibfnamefont {M.-J.}\ \bibnamefont {Hwang}},
  \bibinfo {author} {\bibfnamefont {J.}~\bibnamefont {Casanova}}, \ and\
  \bibinfo {author} {\bibfnamefont {M.~B.}\ \bibnamefont {Plenio}},\
  }\href@noop {} {\bibfield  {journal} {\bibinfo  {journal} {Phys. Rev. Lett.}\
  }\textbf {\bibinfo {volume} {118}},\ \bibinfo {pages} {073001} (\bibinfo
  {year} {2017})}\BibitemShut {NoStop}%
\bibitem [{\citenamefont {Ying}\ \emph {et~al.}(2020)\citenamefont {Ying},
  \citenamefont {Cong},\ and\ \citenamefont {Sun}}]{ying2020quantum}%
  \BibitemOpen
  \bibfield  {author} {\bibinfo {author} {\bibfnamefont {Z.-J.}\ \bibnamefont
  {Ying}}, \bibinfo {author} {\bibfnamefont {L.}~\bibnamefont {Cong}}, \ and\
  \bibinfo {author} {\bibfnamefont {X.-M.}\ \bibnamefont {Sun}},\ }\href@noop
  {} {\bibfield  {journal} {\bibinfo  {journal} {J. Phys. A: Math. Theor.}\
  }\textbf {\bibinfo {volume} {53}},\ \bibinfo {pages} {345301} (\bibinfo
  {year} {2020})}\BibitemShut {NoStop}%
\bibitem [{\citenamefont {Liu}\ \emph {et~al.}(2017)\citenamefont {Liu},
  \citenamefont {Chesi}, \citenamefont {Ying}, \citenamefont {Chen},
  \citenamefont {Luo},\ and\ \citenamefont {Lin}}]{PhysRevLett.119.220601}%
  \BibitemOpen
  \bibfield  {author} {\bibinfo {author} {\bibfnamefont {M.}~\bibnamefont
  {Liu}}, \bibinfo {author} {\bibfnamefont {S.}~\bibnamefont {Chesi}}, \bibinfo
  {author} {\bibfnamefont {Z.-J.}\ \bibnamefont {Ying}}, \bibinfo {author}
  {\bibfnamefont {X.}~\bibnamefont {Chen}}, \bibinfo {author} {\bibfnamefont
  {H.-G.}\ \bibnamefont {Luo}}, \ and\ \bibinfo {author} {\bibfnamefont
  {H.-Q.}\ \bibnamefont {Lin}},\ }\href@noop {} {\bibfield  {journal} {\bibinfo
   {journal} {Phys. Rev. Lett.}\ }\textbf {\bibinfo {volume} {119}},\ \bibinfo
  {pages} {220601} (\bibinfo {year} {2017})}\BibitemShut {NoStop}%
\bibitem [{\citenamefont {Fallas~Padilla}\ \emph {et~al.}(2022)\citenamefont
  {Fallas~Padilla}, \citenamefont {Pu}, \citenamefont {Cheng},\ and\
  \citenamefont {Zhang}}]{fallas2022understanding}%
  \BibitemOpen
  \bibfield  {author} {\bibinfo {author} {\bibfnamefont {D.}~\bibnamefont
  {Fallas~Padilla}}, \bibinfo {author} {\bibfnamefont {H.}~\bibnamefont {Pu}},
  \bibinfo {author} {\bibfnamefont {G.-J.}\ \bibnamefont {Cheng}}, \ and\
  \bibinfo {author} {\bibfnamefont {Y.-Y.}\ \bibnamefont {Zhang}},\ }\href@noop
  {} {\bibfield  {journal} {\bibinfo  {journal} {Phys. Rev. Lett.}\ }\textbf
  {\bibinfo {volume} {129}},\ \bibinfo {pages} {183602} (\bibinfo {year}
  {2022})}\BibitemShut {NoStop}%
\bibitem [{\citenamefont {Ying}\ \emph {et~al.}(2022)\citenamefont {Ying},
  \citenamefont {Felicetti}, \citenamefont {Liu},\ and\ \citenamefont
  {Braak}}]{ying2022critical}%
  \BibitemOpen
  \bibfield  {author} {\bibinfo {author} {\bibfnamefont {Z.-J.}\ \bibnamefont
  {Ying}}, \bibinfo {author} {\bibfnamefont {S.}~\bibnamefont {Felicetti}},
  \bibinfo {author} {\bibfnamefont {G.}~\bibnamefont {Liu}}, \ and\ \bibinfo
  {author} {\bibfnamefont {D.}~\bibnamefont {Braak}},\ }\href@noop {}
  {\bibfield  {journal} {\bibinfo  {journal} {Entropy}\ }\textbf {\bibinfo
  {volume} {24}},\ \bibinfo {pages} {1015} (\bibinfo {year}
  {2022})}\BibitemShut {NoStop}%
\bibitem [{\citenamefont {Chen}\ \emph {et~al.}(2021)\citenamefont {Chen},
  \citenamefont {Wu}, \citenamefont {Jiang}, \citenamefont {L{\"u}},
  \citenamefont {Peng},\ and\ \citenamefont {Du}}]{chen2021experimental}%
  \BibitemOpen
  \bibfield  {author} {\bibinfo {author} {\bibfnamefont {X.}~\bibnamefont
  {Chen}}, \bibinfo {author} {\bibfnamefont {Z.}~\bibnamefont {Wu}}, \bibinfo
  {author} {\bibfnamefont {M.}~\bibnamefont {Jiang}}, \bibinfo {author}
  {\bibfnamefont {X.-Y.}\ \bibnamefont {L{\"u}}}, \bibinfo {author}
  {\bibfnamefont {X.}~\bibnamefont {Peng}}, \ and\ \bibinfo {author}
  {\bibfnamefont {J.}~\bibnamefont {Du}},\ }\href@noop {} {\bibfield  {journal}
  {\bibinfo  {journal} {Nat. Commun.}\ }\textbf {\bibinfo {volume} {12}},\
  \bibinfo {pages} {6281} (\bibinfo {year} {2021})}\BibitemShut {NoStop}%
\bibitem [{\citenamefont {Larson}\ and\ \citenamefont
  {Irish}(2017)}]{Larson_2017}%
  \BibitemOpen
  \bibfield  {author} {\bibinfo {author} {\bibfnamefont {J.}~\bibnamefont
  {Larson}}\ and\ \bibinfo {author} {\bibfnamefont {E.~K.}\ \bibnamefont
  {Irish}},\ }\href@noop {} {\bibfield  {journal} {\bibinfo  {journal} {J.
  Phys. A: Math. Theor.}\ }\textbf {\bibinfo {volume} {50}},\ \bibinfo {pages}
  {174002} (\bibinfo {year} {2017})}\BibitemShut {NoStop}%
\bibitem [{\citenamefont {Ying}\ \emph {et~al.}(2015)\citenamefont {Ying},
  \citenamefont {Liu}, \citenamefont {Luo}, \citenamefont {Lin},\ and\
  \citenamefont {You}}]{PhysRevA.92.053823}%
  \BibitemOpen
  \bibfield  {author} {\bibinfo {author} {\bibfnamefont {Z.-J.}\ \bibnamefont
  {Ying}}, \bibinfo {author} {\bibfnamefont {M.}~\bibnamefont {Liu}}, \bibinfo
  {author} {\bibfnamefont {H.-G.}\ \bibnamefont {Luo}}, \bibinfo {author}
  {\bibfnamefont {H.-Q.}\ \bibnamefont {Lin}}, \ and\ \bibinfo {author}
  {\bibfnamefont {J.}~\bibnamefont {You}},\ }\href@noop {} {\bibfield
  {journal} {\bibinfo  {journal} {Phys. Rev. A}\ }\textbf {\bibinfo {volume}
  {92}},\ \bibinfo {pages} {053823} (\bibinfo {year} {2015})}\BibitemShut
  {NoStop}%
\bibitem [{\citenamefont {Lu}\ and\ \citenamefont {Zhang}(2024)}]{Lu_2024}%
  \BibitemOpen
  \bibfield  {author} {\bibinfo {author} {\bibfnamefont {Y.-q.}\ \bibnamefont
  {Lu}}\ and\ \bibinfo {author} {\bibfnamefont {Y.-Y.}\ \bibnamefont {Zhang}},\
  }\href@noop {} {\bibfield  {journal} {\bibinfo  {journal} {New J. Phys.}\
  }\textbf {\bibinfo {volume} {26}},\ \bibinfo {pages} {063010} (\bibinfo
  {year} {2024})}\BibitemShut {NoStop}%
\bibitem [{\citenamefont {Grimaudo}\ \emph
  {et~al.}(2023{\natexlab{a}})\citenamefont {Grimaudo}, \citenamefont
  {Messina}, \citenamefont {Sergi}, \citenamefont {Solano},\ and\ \citenamefont
  {Valenti}}]{grimaudo2023thermodynamic}%
  \BibitemOpen
  \bibfield  {author} {\bibinfo {author} {\bibfnamefont {R.}~\bibnamefont
  {Grimaudo}}, \bibinfo {author} {\bibfnamefont {A.}~\bibnamefont {Messina}},
  \bibinfo {author} {\bibfnamefont {A.}~\bibnamefont {Sergi}}, \bibinfo
  {author} {\bibfnamefont {E.}~\bibnamefont {Solano}}, \ and\ \bibinfo {author}
  {\bibfnamefont {D.}~\bibnamefont {Valenti}},\ }\href@noop {} {\bibfield
  {journal} {\bibinfo  {journal} {arXiv:2310.19595}\ } (\bibinfo {year}
  {2023}{\natexlab{a}})}\BibitemShut {NoStop}%
\bibitem [{\citenamefont {Braak}(2011)}]{PhysRevLett.107.100401}%
  \BibitemOpen
  \bibfield  {author} {\bibinfo {author} {\bibfnamefont {D.}~\bibnamefont
  {Braak}},\ }\href@noop {} {\bibfield  {journal} {\bibinfo  {journal} {Phys.
  Rev. Lett.}\ }\textbf {\bibinfo {volume} {107}},\ \bibinfo {pages} {100401}
  (\bibinfo {year} {2011})}\BibitemShut {NoStop}%
\bibitem [{\citenamefont {Grimaudo}\ \emph
  {et~al.}(2023{\natexlab{b}})\citenamefont {Grimaudo}, \citenamefont
  {de~Castro}, \citenamefont {Messina}, \citenamefont {Solano},\ and\
  \citenamefont {Valenti}}]{PhysRevLett.130.043602}%
  \BibitemOpen
  \bibfield  {author} {\bibinfo {author} {\bibfnamefont {R.}~\bibnamefont
  {Grimaudo}}, \bibinfo {author} {\bibfnamefont {A.~M.}\ \bibnamefont
  {de~Castro}}, \bibinfo {author} {\bibfnamefont {A.}~\bibnamefont {Messina}},
  \bibinfo {author} {\bibfnamefont {E.}~\bibnamefont {Solano}}, \ and\ \bibinfo
  {author} {\bibfnamefont {D.}~\bibnamefont {Valenti}},\ }\href@noop {}
  {\bibfield  {journal} {\bibinfo  {journal} {Phys. Rev. Lett.}\ }\textbf
  {\bibinfo {volume} {130}},\ \bibinfo {pages} {043602} (\bibinfo {year}
  {2023}{\natexlab{b}})}\BibitemShut {NoStop}%
\bibitem [{\citenamefont {Schneider}\ \emph {et~al.}(2012)\citenamefont
  {Schneider}, \citenamefont {Porras},\ and\ \citenamefont
  {Schaetz}}]{Schneider_2012}%
  \BibitemOpen
  \bibfield  {author} {\bibinfo {author} {\bibfnamefont {C.}~\bibnamefont
  {Schneider}}, \bibinfo {author} {\bibfnamefont {D.}~\bibnamefont {Porras}}, \
  and\ \bibinfo {author} {\bibfnamefont {T.}~\bibnamefont {Schaetz}},\
  }\href@noop {} {\bibfield  {journal} {\bibinfo  {journal} {Rep. Prog. Phys.}\
  }\textbf {\bibinfo {volume} {75}},\ \bibinfo {pages} {024401} (\bibinfo
  {year} {2012})}\BibitemShut {NoStop}%
\bibitem [{\citenamefont {Lemmer}\ \emph {et~al.}(2018)\citenamefont {Lemmer},
  \citenamefont {Cormick}, \citenamefont {Tamascelli}, \citenamefont {Schaetz},
  \citenamefont {Huelga},\ and\ \citenamefont {Plenio}}]{Lemmer_2018}%
  \BibitemOpen
  \bibfield  {author} {\bibinfo {author} {\bibfnamefont {A.}~\bibnamefont
  {Lemmer}}, \bibinfo {author} {\bibfnamefont {C.}~\bibnamefont {Cormick}},
  \bibinfo {author} {\bibfnamefont {D.}~\bibnamefont {Tamascelli}}, \bibinfo
  {author} {\bibfnamefont {T.}~\bibnamefont {Schaetz}}, \bibinfo {author}
  {\bibfnamefont {S.~F.}\ \bibnamefont {Huelga}}, \ and\ \bibinfo {author}
  {\bibfnamefont {M.~B.}\ \bibnamefont {Plenio}},\ }\href@noop {} {\bibfield
  {journal} {\bibinfo  {journal} {New J. Phys.}\ }\textbf {\bibinfo {volume}
  {20}},\ \bibinfo {pages} {073002} (\bibinfo {year} {2018})}\BibitemShut
  {NoStop}%
\end{thebibliography}%
\end{document}